\documentclass[prb,twocolumn,superscriptaddress,longbibliography,floatfix]{revtex4-2}
\usepackage{graphicx,amsfonts,amssymb,amsmath,xspace,dsfont}
\usepackage{bbm}
\usepackage[colorlinks=true,citecolor=BurntOrange,linkcolor=red,urlcolor=Bittersweet]{hyperref}
\usepackage{multirow}
\usepackage{color}
\usepackage{wasysym}
\usepackage[dvipsnames]{xcolor}
\definecolor{g}{rgb}{.1,0.4,.1} 
\definecolor{b}{rgb}{0,0.2,1}
\definecolor{rouge}{rgb}{0.82,0.,0.}
\definecolor{vert}{rgb}{0.,0.82,0.}
\definecolor{orange}{rgb}{1,0.5,0.}
\definecolor{bleu}{rgb}{0.,0.,0.82}
\definecolor{m}{rgb}{0.82,0.,0.82}
\definecolor{vert2}{rgb}{0.,0.5,0.}
\definecolor{rougeclair}{rgb}{1.0,0.7,0.7}

\newcommand{\be}{\begin{equation}}
\newcommand{\ee}{\end{equation}}
\newcommand{\beqn}{\begin{eqnarray}}
\newcommand{\eeqn}{\end{eqnarray}}

\usepackage{bm}

\newcommand{\ket}[1]{|#1\rangle}

\begin{document}


\title{Kitaev-Heisenberg model on the square-hexagon-dodecagon lattice}

\author{Saeed Barari}
\email{saeedbarari@iasbs.ac.ir}
\affiliation{Department of Physics, Institute for Advanced Studies in Basic Sciences (IASBS), Zanjan 45137-66731, Iran}

\author{Yasir Iqbal}
\email{yiqbal@physics.iitm.ac.in}
\affiliation{Department of Physics, Indian Institute of Technology Madras, Chennai 600036, India}

\author{Ganapathy Baskaran}
\email{baskaran@imsc.res.in}
\affiliation{The Institute of Mathematical Sciences, CIT Campus, Chennai 600 113, India}
\affiliation{Department of Physics, Indian Institute of Technology Madras, Chennai 600036, India}
\affiliation{Perimeter Institute for Theoretical Physics, Waterloo, ON N2L 2Y5, Canada}

\author{Saeed S. Jahromi}
\email{saeed.jahromi@iasbs.ac.ir}
\affiliation{Department of Physics, Institute for Advanced Studies in Basic Sciences (IASBS), Zanjan 45137-66731, Iran}
\affiliation{Department of Physics, Indian Institute of Technology Madras, Chennai 600036, India}
\affiliation{Donostia International Physics Center, Paseo Manuel de Lardizabal 4, E-20018 San Sebasti\'an, Spain}

\author{Chunxiao Liu}
\email{chunxiao.liu@universite-paris-saclay.fr}
\affiliation{Universit\'e Paris-Saclay, CNRS, Laboratoire de Physique des Solides, 91405 Orsay, France}

\author{Julien Vidal}
\email{julien.vidal@sorbonne-universite.fr}
\affiliation{Sorbonne Universit\'e, CNRS, Laboratoire de Physique Th\'eorique de la Mati\`ere Condens\'ee, LPTMC, F-75005 Paris, France}
\affiliation{Department of Physics, Indian Institute of Technology Madras, Chennai 600036, India}

\begin{abstract}
We study the spin-$1/2$ Kitaev-Heisenberg model on the square-hexagon-dodecagon lattice with a symmetry-preserving tri-coloring of the Kitaev exchanges. We focus on two cases: (i) the pure Kitaev model with anisotropic exchanges and (ii) the Kitaev-Heisenberg model with isotropic Kitaev exchange. In the absence of Heisenberg interactions, we use the standard mapping of the Kitaev model onto a quadratic Majorana-fermion problem in a $\mathbb{Z}_2$ gauge field. We find a gapped energy spectrum throughout the positive-coupling phase diagram, except at a single point where the twofold-degenerate energy bands form a Dirac cone. This unique gapped region corresponds to a toric code phase. However, depending on the couplings, vortex excitations on square, hexagonal, and dodecagonal plaquettes realize three distinct relative assignments to the $e$ and $m$ anyons. 
We determine these relative assignments from the physical fermion parity of the vortex sectors and locate their boundaries by tracking zero crossings of vortex-bound Majorana levels.
For the isotropic Kitaev-Heisenberg model, we combine graph-based projected entangled-pair-state calculations, exact diagonalizations, and linear spin-wave theory to study the entire phase diagram. Apart from the toric code topological phases that are robust around the Kitaev limits, we obtain four collinear magnetically ordered phases that can be characterized by their local and relative orderings of hexagonal plaquettes. We also identify a special point at which the ground state is a product state.
\end{abstract}

\maketitle

\section{Introduction}
\label{sec:intro}

Quantum spin liquids (QSLs) are a paradigmatic class of exotic phases of matter  in which strong fluctuations prevent conventional magnetic order down to zero temperature.  They are characterized by fractionalized excitations and emergent gauge fields. Originally proposed in the context of frustrated quantum magnets, QSLs have attracted sustained interest due to their deep connections to topological order, and potential applications in fault-tolerant quantum computation. 

A particularly influential and rare example of a model featuring QSLs is provided by the Kitaev model~\cite{Kitaev06}. Originally defined on the honeycomb lattice, this interacting \mbox{spin-1/2} model, which is based on bond-dependent Ising interactions, can be solved exactly by fractionalizing spins into itinerant Majorana fermions coupled to a static $\mathbb{Z}_2$ gauge field. The solvability of the Kitaev model is not restricted to the honeycomb lattice. It extends to any tricoordinated graph that admits a three-edge coloring.  Thus, this model has been studied in several different two-dimensional geometries including amorphous systems~\cite{Cassella23,Grushin23}, and, more recently, hyperbolic lattices~\cite{Mosseri25,Dusel25,Lenggenhager25,Vidal25}), revealing a wide variety of phases. Importantly, for a given geometry, distinct colorings also produce different phase diagrams (see, e.g., Refs~\cite{Kamfor10,Quinn15} for such a study in the honeycomb lattice). This raises the question of which properties of a Kitaev spin liquid are generic consequences of tricoordination and bond-dependent exchange, and which depend on the geometry and coloring of the lattice~?  Many works have shown that the answer is sensitive to both.

\begin{figure}
\centerline{\includegraphics[width=0.9\columnwidth]{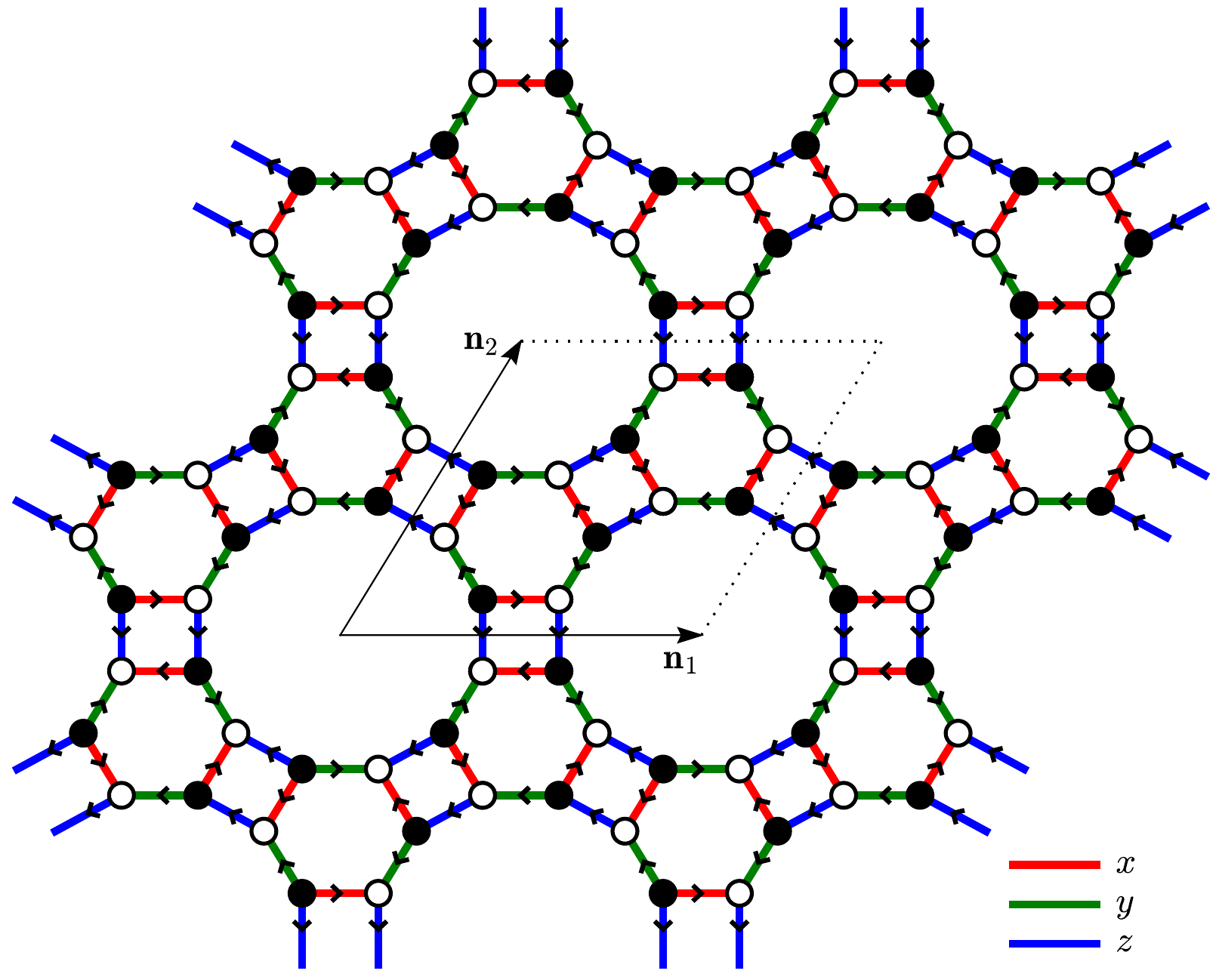}}
\caption{SHD lattice and tri-coloring used in this work.  The red,
green, and blue bonds denote the three Kitaev bond types.  The arrows
specify one periodic choice of the $\mathbb Z_2$ gauge variables
$u_{jk}$ realizing the ground-state flux sector of the Kitaev model at
$J=0$.  The primitive vectors $\mathbf n_1$ and $\mathbf n_2$ define
the twelve-site unit cell used for the Majorana Bloch Hamiltonian and
for the tensor-network calculations.}
\label{fig:kekule_coloring}
\end{figure}

Archimedean lattices provide a controlled setting in which to study this geometry dependence.  These lattices are vertex-transitive (i.e., all sites are equivalent) and are built from regular polygons. They only differ in their elementary loops and unit-cell structures.  Of the eleven Archimedean lattices, four are tricoordinated allowing for the consistent assignment of the three distinct types of bond required by the Kitaev construction: the \mbox{honeycomb lattice $(6^3)$~\cite{Kitaev06}}, the triangle-dodecagon lattice $(3,12^2)$~\cite{Yao07,Dusuel08_2}, the square-octagon lattice $(4,8^2)$~\cite{Baskaran09}, and the square-hexagon-dodecagon lattices (SHD) $(4,6,12$).  The SHD lattice is the main topic of this work.  It contains three even-length plaquettes of different types and the three-edge (tri-)coloring considered here requires a twelve-site unit cell (see Fig.~\ref{fig:kekule_coloring}).  Importantly, this coloring is not invariant under arbitrary permutations of the three bond labels. 

In this work, we study the Kitaev-Heisenberg Hamiltonian
\begin{equation}
H
=
\sum_{\alpha=x,y,z} K_\alpha \sum_{\langle i,j\rangle_\alpha}
 S_i^\alpha S_j^\alpha
+
J\sum_{\langle i,j\rangle}\mathbf S_i\cdot\mathbf S_j,
\label{eq:kh}
\end{equation}
on the SHD lattice with the tri-coloring shown in Fig.~\ref{fig:kekule_coloring} and  we focus on two cases.  First, we set $J=0$ and study the pure Kitaev model with independently varying exchanges $K_x$, $K_y$, and $K_z$.  Second, we set \mbox{$K_x=K_y=K_z=K$} and study the isotropic Kitaev-Heisenberg model as a function of $K$ and $J$.

For the pure Kitaev model, we solve the Majorana problem in the ground-state flux sector as originally proposed by Kitaev~\cite{Kitaev06}.  The required flux pattern can be represented by a periodic gauge choice within the same twelve-site unit cell as the spin model.  The corresponding Majorana Hamiltonian has projective lattice symmetries.  At each momentum its six positive-energy eigenvalues form three degenerate doublets.  We find that the Majorana matter spectrum in the triangular phase diagram in anisotropic-coupling space, defined by $K_x,K_y,K_z \geqslant 0$ and $K_x+K_y+K_z=1$, is gapped everywhere except at one point (see Fig.~\ref{fig:Kitaev_PD}).  The isotropic point is gapped, and its time-reversal-symmetric Majorana bands have vanishing Chern number. Together with the isolated-dimer assignments discussed below, these results identify the connected gapped region with the Abelian toric code phase. Actually, this phase can be further split into three regions with different relative assignments of plaquette vortices to the $e$ and $m$ anyons (see Fig.~\ref{fig:anyon_regions}). This identification is straightforward in the isolated-dimer limits where one of the couplings dominates. Following the physical-parity criterion of Ref.~\cite{Chern26}, we then extend the assignments to generic couplings.  Crossing a boundary between two regions changes the species assigned to one plaquette class through a zero crossing of a vortex-bound Majorana level, while the vortex-free spectrum remains gapped.  

We then restrict Eq.~\eqref{eq:kh} to $K_x=K_y=K_z=K$ and study the $(K,J)$ phase diagram.  This model allows us to examine how the gapped Kitaev spin liquid is destabilized by the simplest isotropic exchange interaction.  On the honeycomb lattice, the Kitaev-Heisenberg model contains stripy and zigzag phases between the spin-liquid and conventional magnetic regimes \cite{Chaloupka10,Chaloupka13,Chaloupka15}.  On the SHD lattice, the larger unit cell leads to a different organization of magnetic order: the relevant distinction is between the local order on a hexagon and the relative ordering between neighboring hexagons. Using graph-based projected entangled-pair-state (gPEPS) calculations, benchmarked against exact diagonalization, we obtain six different phases (see Fig.~\ref{fig:phase_diagram}).  Linear spin-wave theory provides an additional approach within the stable interiors of the ordered phases.  Two narrow quantum spin-liquid regions surround the two exactly solved, gapped Kitaev limits.  Their continuity with these limits identifies them with the same toric code phase.  The remaining four regions are magnetically ordered.  The conventional AFM$_+$ and FM$_+$ phases are connected to the antiferromagnetic and ferromagnetic Heisenberg limits, respectively.  Between these conventional orders and the Kitaev spin liquids we find two intermediate phases, AFM$_-$ and FM$_-$.  Compared to the AFM$_+$ and FM$_+$ phases, the local order on each hexagon is retained, but the relative ordering between neighboring hexagons is staggered.  We also identify an exact product-state point at $(K,J)=(-2,1)$, where the Hamiltonian is frustration free and the ground state is of AFM$_-$ type.

The remainder of the paper follows these two questions in sequence. Section~\ref{sec:Kitaev_SHD} focuses on the pure Kitaev Hamiltonian with anisotropic exchanges.  We determine the ground-state flux sector and the projective symmetries of the model, solve the Majorana spectrum to locate the gapless point in the phase diagram, connect the surrounding gapped region to the anisotropic toric code limits, and track the relative anyon assignments. Section~\ref{sec:Kitaev_Heisenberg_SHD} turns to the effect of Heisenberg exchange in the isotropic Kitaev model.  The classical orders provide the framework for interpreting the gPEPS phase diagram, while exact diagonalization and stable linear spin-wave theory give complementary benchmarks; the section concludes with the exact frustration-free ground state at $(K,J)=(-2,1)$. Section~\ref{sec:conclude} discusses what these results imply for the role of lattice geometry in Kitaev materials and identifies questions for future studies.  
Appendices~\ref{app:gpeps}--\ref{app:exact_solvable_point} collect the technical details of the tensor-network, exact-diagonalization, Majorana and spin-wave calculations, as well as information about an exactly solvable point.

\section{Pure Kitaev model}
\label{sec:Kitaev_SHD}

We first analyze the pure Kitaev model, obtained by setting $J=0$ in Eq.~\eqref{eq:kh}.  We recall some basic properties of the model and present a gauge choice that achieves the ground-state flux sector. Finally, we compute the energy spectrum and the ground-state phase diagram.

\subsection{Generalities}
\label{sec:kitaev_generalities}

\mbox{At $J=0$}, Eq.~\eqref{eq:kh} reduces to the Kitaev Hamiltonian
\begin{equation}
H_{\rm K}
=\sum_{\alpha=x,y,z} K_\alpha \sum_{\langle i,j\rangle_\alpha}
 S_i^\alpha S_j^\alpha,
\label{eq:ham0}
\end{equation}
where the label $\alpha$ is fixed by the three-edge coloring in Fig.~\ref{fig:kekule_coloring}. Here, we use $S_i^\alpha=\sigma_i^\alpha/2$, with $\sigma_i^\alpha$ a Pauli
matrix.  

For any plaquette $p$ with oriented boundary $\partial p$,  we define
%
%
\begin{equation}
W_p
=
\prod_{(j,k)\in\partial p}
\sigma_j^{\alpha(jk)}
\sigma_k^{\alpha(jk)} ,
\label{eq:wpdef}
\end{equation}
%
%
where $\alpha(jk)$ denotes the color of the bond $(j,k)$ and the product follows a fixed orientation (e.g., clockwise).   One can readily verify that the operator is a conserved quantity, i.e., $[H_{\rm K},W_p]=0$. Furthermore, one has  $[W_p,W_{p'}]=0$ for any pair of plaquettes $p$ and $p'$, so that the Hamiltonian can be block-diagonalized into sectors labeled by the eigenvalues of the set ${W_p}$. For a fixed configuration of the $W_p$'s, the Hamiltonian~\eqref{eq:ham0} can then be written as
%
%
\be 
H_{\rm K}(\{u\})=-\frac{\mathrm{i}}{4} \sum_{j,k} A_{jk} c_j c_k,
\label{eq:ham_Majo}
\ee  
%
%
where the sum runs over all lattice sites and $c_j$ denotes a Majorana fermion operator acting on site $j$. As originally shown by Kitaev, this mapping enlarges the Hilbert space; therefore, a projection onto the physical subspace must be performed to recover the physical spin states~\cite{Kitaev06}. The matrix $A$ is a real skew-symmetric matrix whose elements depend on $\mathbb{Z}_2$ gauge variables \mbox{$u_{jk}=-u_{kj}=\pm1$} defined on each bond of the lattice. More precisely, if sites $j$ and $k$ are connected by a bond of type $\alpha$, the corresponding matrix element is given by $A_{jk}=K_{\alpha}u_{jk}/2$ (see Ref.~\cite{Kitaev06} for a detailed derivation). 
In this representation, the eigenvalue of the plaquette operator $W_p$, interpreted as the flux in the plaquette $p$, can be expressed as
%
%
\be
w_p=(-{\rm i})^q \prod_{(j,k)\in {\partial p}} u_{jk},
\label{eq:fluxdef}
\ee
%
%
where $q$ denotes the number of edges of the plaquette $p$, and the product runs over all links $(j,k)$ along the oriented boundary $\partial p$ of the plaquette~\cite{Kitaev06}. Thus, for any fixed configuration of the gauge variables $u_{jk}$, the Hamiltonian reduces to a quadratic form of Majorana fermions. On a closed surface, because every link appears
twice when multiplying over all plaquettes, fluxes satisfy the global constraint $\prod_p w_p=1$, or equivalently $\prod_p W_p=\mathds{1}$. This condition reflects the well-known total flux quantization on a closed oriented surface~\cite{Dirac31}.

The remaining question is which flux sector contains the ground state. For strictly positive couplings $K_\alpha$, the infinite SHD graph is planar and bipartite, and the colored bond magnitudes are invariant under the reflection axes that bisect its elementary even plaquettes.  Lieb's flux-phase argument therefore selects~\cite{Lieb93,Lieb94}
\begin{equation}
w_p^{\rm g.s.}
=
-(\pm {\rm i})^{q} ,
\label{eq:fluxmin}
\end{equation}
where $q$ is the number of edges of plaquette $p$.  On the SHD lattice
this gives
\begin{equation}
w_p^{\rm g.s.}
=
\begin{cases}
-1, & \text{ if $p$ is a square or a dodecagon},\\
+1, &\text{ if $p$ is a hexagon}.
\end{cases}
\label{eq:SHD_flux_pattern}
\end{equation}
For nonzero couplings of arbitrary sign, let $s_\alpha=\operatorname{sgn}K_\alpha$ and absorb these signs into the effective links $\widetilde u_{jk}=s_\alpha u_{jk}$ on an $\alpha$ bond. The Majorana hopping problem then has positive magnitudes $K_\alpha$, so Lieb's result applies to the effective plaquette flux.  In terms of the physical spin flux, Eq.~\eqref{eq:fluxmin} acquires the factor $\prod_{(j,k)_\alpha\in\partial p}s_\alpha$.  Equation~\eqref{eq:SHD_flux_pattern} is therefore unchanged whenever this product is $+1$, including both signs of the isotropic Kitaev coupling because all SHD plaquettes have even length. On a finite torus, Eq.~\eqref{eq:SHD_flux_pattern} fixes the local plaquette fluxes but not the two independent Wilson loops.

\subsection{Coloring and gauge choice}
\label{sec:coloring_gauge}

We now discuss the coloring in Fig.~\ref{fig:kekule_coloring}.  It preserves the point-group operations used below and has a twelve-site unit cell containing two hexagons. 
For a given coloring of the lattice (i.e., a fixed assignment of the $x$, $y$, and $z$ bonds), the next step is to determine a gauge configuration that realizes the desired flux pattern. Interestingly, for the present system, it is possible to choose a set of gauge variables $u_{jk}$ that reproduces the ground-state flux sector discussed above [see Eq.~\eqref{eq:fluxmin}] and that has the same periodicity as the twelve-site unit cell. More generally, for any periodic lattice, such a gauge choice exists if and only if
%
%
\be
\prod_{p \: \in \textrm{~unit cell}} w_p^{{\rm g.s.}}=+1. 
\ee
%
%

The SHD unit cell with our coloring contains two hexagons, three squares, and one dodecagon, and the Lieb pattern indeed has plaquette-flux product $(+1)^2(-1)^3(-1)=+1$. The gauge-fixed Majorana Hamiltonian consequently retains the same twelve-site primitive cell as the spin model.  Its $12\times12$ Bloch matrix $\mathcal{H}(\mathbf{k})$ is given explicitly in
Appendix~\ref{app:free_majorana_hamiltonian}.
\subsection{Symmetry and projective symmetry}
\label{sec:projective_symmetry}

The model \eqref{eq:kh}
with the tri-coloring given in
Fig.~\ref{fig:kekule_coloring} respects the full symmetry of the SHD lattice, given by the 
wallpaper group
$p6m\simeq\mathbb Z^2\rtimes D_6$.  The Kitaev
interaction also preserves the global spin rotation group
$\mathbb Z_2^2$, generated by the product of all Pauli matrices $X=\prod_i \sigma_i^x$, $Y=\prod_i \sigma_i^y$, and $Z=\prod_i\sigma_i^z$, as well as the physical time-reversal
$\mathcal T$. The complete symmetry is
therefore
\begin{equation}
G=(p6m\times \mathbb Z_2^2) \rtimes \mathbb{Z}_2^{\mathcal{T}}.
\end{equation}
Below we choose the generators of $p6m$ to be (a) $C_3$, the counterclockwise
$2\pi/3$ rotation about the center of the lower-left $A$ hexagon, (b)
$C_2$, the $\pi$ rotation about the center of the reference square,
and (c) \mbox{$M_x:(x,y)\mapsto(-x,y)$}, the reflection about the vertical mirror axis. 


We now study the symmetry action on the Majorana Hamiltonian \eqref{eq:ham_Majo}. Denote the Majorana Bloch Hamiltonian matrix as
$\mathcal H(\mathbf k)$, whose explicit form is given in
Appendix~\ref{app:free_majorana_hamiltonian}. The gauge chosen in Fig.~\ref{fig:kekule_coloring} can have a nontrivial effect:  a spatial operation
can preserve all plaquette fluxes while changing the link representative.
Its action in Majorana space must then be followed by a local gauge
transformation that restores the gauge, hence the Majorana carries a \emph{projective representation} of the group $G$ (see Ref.~\cite{Fuchs20} for a related discussion in the honeycomb lattice). Explicitly, we have
\begin{subequations}
\label{eq:projective_symmetry_relations}
\begin{align}
 U_{C_3}\,
\mathcal H(\mathbf k)\,
 U_{C_3}^{\dagger}
&=
\mathcal H\!\left[
C_3^{-1}(\mathbf k+\mathbf b)
\right],
\label{eq:C3_projective}
\\
 U_{C_2}\,
\mathcal H(\mathbf k)\,
 U_{C_2}^{\dagger}
&=
\mathcal H(-\mathbf k),
\label{eq:C2_projective}
\\
 U_{M_x}\,
\mathcal H(\mathbf k)\,
 U_{M_x}^{\dagger}
&=
\mathcal H\!\left(M_x\mathbf k+\mathbf b\right),
\label{eq:M_projective}
\\
\mathcal{K}\mathcal H(\mathbf k)\mathcal{K}^{-1} \equiv \mathcal H^*(\mathbf k)
&=
-\mathcal H(-\mathbf k),
\label{eq:majorana_reality_symmetry}
\\
\Gamma\mathcal H(\mathbf k)\Gamma^{-1}
&=-\mathcal H(\mathbf k).
\label{eq:majorana_chiral_symmetry}
\end{align}
\end{subequations}
Here $ U_S \equiv U_S(\mathbf k)$ is the unitary matrix associated
with the spatial symmetry $S$, which combines the site permutation, the
gauge-restoring signs, and any Bloch phases generated by translations
across unit-cell boundaries.  Explicit form of the matrices
$U_S$ and $\Gamma$
are given in Appendix~\ref{app:free_majorana_hamiltonian}.  The momentum
shift $\mathbf b$ appearing in Eqs.~\eqref{eq:C3_projective} and
\eqref{eq:M_projective} satisfies $\mathbf{b}\cdot\mathbf{n}_1=0$ and $\mathbf{b}\cdot\mathbf{n}_2=\pi$ and is part of the projective representation. Equation~\eqref{eq:majorana_reality_symmetry} accounts for the  Majorana reality condition, where the operator $\mathcal{K}$ denotes complex conjugation, and Eq.~\eqref{eq:majorana_chiral_symmetry} defines the chiral symmetry of the Majorana Hamiltonian. The product $\Gamma\mathcal K$
gives rise to the physical time-reversal symmetry $\mathcal{T}$.  Both $C_2$ and
$\mathcal{T}$ reverse momentum, so their composition $\Theta:=C_2 \circ \mathcal{T}$ acts within the Bloch space at fixed $\mathbf k$; since the operation $\Theta$ commutes
with $\mathcal H(\mathbf k)$ and squares to $-1$, it produces Kramers doublets at any momentum, and results in a twofold degeneracy for every Majorana band throughout the Brillouin zone.


\subsection{Spectrum and phase diagram}
\label{sec:kitaev_spectrum}

\begin{figure}[t]
\centering
\includegraphics[width=0.98\columnwidth]{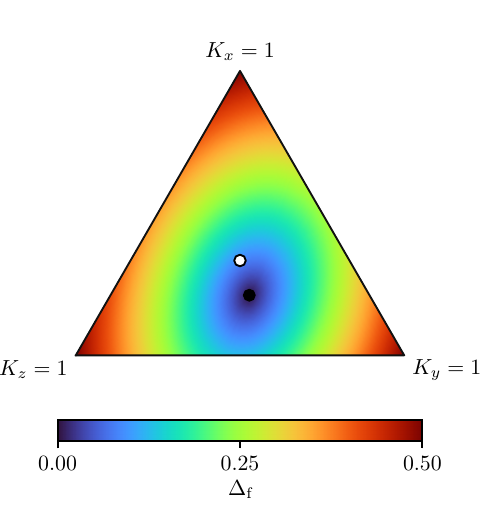}
\caption{Fermionic gap $\Delta_{\rm f}$, defined as the minimum positive
Majorana excitation energy, in the ground-state flux sector of the SHD
Kitaev model.  The triangular anisotropic coupling phase diagram is
defined by $K_x+K_y+K_z=1$ with $K_\alpha\geqslant0$.  The gap is finite
except at the point given in Eq.~\eqref{eq:gapless} (black dot).  The white dot marks the isotropic
point \mbox{$K_x=K_y=K_z$}, which is gapped and is used as the Kitaev limit in
the Kitaev-Heisenberg analysis of
Sec.~\ref{sec:Kitaev_Heisenberg_SHD}.}
\label{fig:Kitaev_PD}
\end{figure}

In the ground-state flux sector, Majorana reality pairs positive and
negative energies, while the symmetry $\Theta$ produces a twofold degeneracy at
each momentum.  The twelve eigenvalues therefore comprise three
positive doublets and their negative partners.  Denoting the physical
positive Majorana energies by $\varepsilon_n(\mathbf k)$, their
characteristic equation is $P(2\varepsilon_n,\mathbf k)=0$, where
\begin{equation}
\begin{aligned}
P(x,\mathbf k)
={}&
x^6
-3S_2x^4
+3x^2\left(S_4+2K_x^2K_z^2\right)
\\
&+3x^2K_y^2\left(K_x^2+K_z^2\right)
\\
&-S_6
-2K_x^3K_y^3
-3\left(K_x^2K_z^4+K_x^4K_z^2\right)
\\
&-2K_xK_y^3K_z^2\,\mathcal{E}(\mathbf k).
\end{aligned}
\label{eq:charpoly}
\end{equation}
Here $S_n=K_x^n+K_y^n+K_z^n$, and all momentum dependence enters through
\begin{equation}
\mathcal{E}(\mathbf k)
:=
\cos(\mathbf k\cdot\mathbf n_1)
+
\cos(\mathbf k\cdot\mathbf n_2)
-
\cos[\mathbf k\cdot(\mathbf n_1-\mathbf n_2)] .
\label{eq:Phi_definition}
\end{equation}
The polynomial $P$ shows that the three couplings are not on equal footing, reflecting the fact that the coloring in Fig.~\ref{fig:kekule_coloring} is not
invariant under arbitrary permutations of the labels $x$, $y$, and $z$.
Consequently, the fermionic gap is not symmetric under all permutations of $(K_x,K_y,K_z)$, as shown in Fig.~\ref{fig:Kitaev_PD}.

\begin{figure}[t]
    \centering
    \includegraphics[width=0.6\columnwidth]{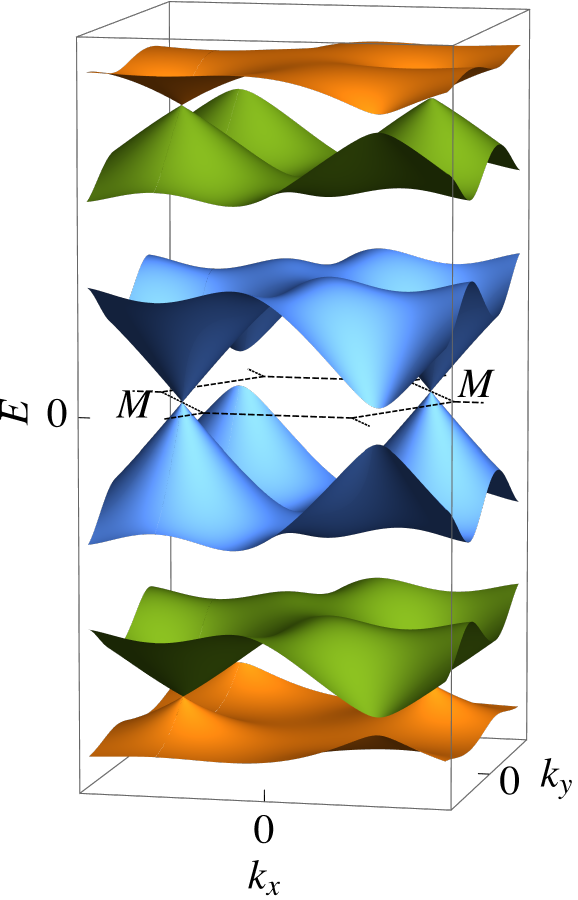}
    \caption{Majorana band structure in the ground-state flux sector at the parameter point \eqref{eq:gapless}. The three positive-energy doublets and their negative-energy Majorana partners are shown. The middle bands touch linearly at zero energy at $\mathrm{M} =(\pi,\frac{\pi}{\sqrt{3}})$ modulo reciprocal lattice vectors (assuming $|\mathbf{n}_1|=|\mathbf{n}_2|=1$), forming a twofold-degenerate Dirac cone. The dashed black contour marks the first Brillouin zone.}
\label{fig:majorana_gapless_dispersion}
\end{figure}

The gap closes when one of the roots of Eq.~\eqref{eq:charpoly}
vanishes. The equation $P(0,\mathbf{k})=0$ has a solution only 
when the couplings satisfy
\begin{equation}
K_x
=
\frac{K_y}{2}
=
\frac{K_z}{\sqrt{3}},
\label{eq:gapless}
\end{equation}
and the momentum solution $\mathbf{k}$ occurs at $\mathrm{M} =(\pi,\frac{\pi}{\sqrt{3}})$ point (assuming $|\mathbf{n}_1|=|\mathbf{n}_2|=1$) of the Brillouin zone, modulo reciprocal lattice vectors. This means that the phase diagram is generically gapped, except at this \emph{isolated} point \eqref{eq:gapless} shown as a black dot in Fig.~\ref{fig:Kitaev_PD}.  Expanding the energy in the vicinity of $\mathrm{M}$ gives a linear dispersion, hence the band structure contains a (twofold-degenerate) Dirac cone. The band structure at the gapless point \eqref{eq:gapless} is shown in Fig.~\ref{fig:majorana_gapless_dispersion}.



An isolated gapless point is not generic among the phase diagram of  Kitaev models in two-dimensional tricoordinated systems. To our knowledge, the only other known example so far is the Kekul\'e-colored honeycomb introduced in Ref.~\cite{Kamfor10} (see also Ref.~\cite{Quinn15} for a related study). However, contrary to this latter case, for the SHD model, the isolated gapless point is not located  at the isotropic coupling, as fixed by Eq.~\eqref{eq:gapless}. This position reflects the absence of full permutation symmetry among the three bond labels. 
Away from this isolated point, the phase diagram consists of  a time-reversal-symmetric gapped region. Hence, the Chern number vanishes throughout this region and according to Kitaev's sixteenfold classification~\cite{Kitaev06}, one has  a topological phase with toric code anyons (see below for more details).  

At the isotropic point $K_x=K_y=K_z=K$ (white dot in Fig.~\ref{fig:Kitaev_PD}), which is the starting point for the analysis of the Kitaev-Heisenberg model in Sec.~\ref{sec:Kitaev_Heisenberg_SHD}, the ground-state energy per site is
\begin{equation}
\varepsilon_0
=
\lim_{N_{\rm uc}\rightarrow\infty}
-\frac{1}{24N_{\rm uc}}
\sum_{\mathbf k}\sum_{n=1}^{6}\varepsilon_n(\mathbf k)
\simeq
-0.2021825\,|K| ,
\label{eq:gse_exact}
\end{equation}
where $N_{\rm uc}$ is the number of twelve-site unit cells.  The corresponding fermionic gap
is
\begin{equation}
\Delta_{\rm f}
=
\frac{1}{2}\sqrt{2-\sqrt3}\,|K|
\simeq
0.258819\,|K| .
\label{eq:gapiso}
\end{equation}


\subsection{Relative anyon assignments}
\label{sec:anyon_assignment}

When one of the couplings dominates, i.e., in the so-called isolated-dimer limit, the Majorana energy spectrum is completely flat. Since the Bloch Hamiltonian $\mathcal H(\mathbf k)$ is time-reversal invariant, the corresponding Chern number is $\nu=0$. According to the sixteenfold-way classification proposed by Kitaev, this corresponds to a toric code phase, which for the SHD lattice, spans the entire phase diagram (except the gapless point). 

In this framework, a complex matter fermion represents the toric code anyon $\psi$,
while plaquette (vortex) excitations carry the two bosonic species $e$ and $m$, which obey mutual semionic statistics. Together with the vacuum $\mathbf{1}$, these are the four anyons of the toric code phase. 

The goal of this section is to determine which plaquettes host $e$ and $m$ excitations. To this aim, we follow the work by Chern et al.~\cite{Chern26}. In the isolated-dimer limit, two plaquettes separated by a strong link are identified as being of the same type.  The $e$ and $m$ particles are distinct anyons exchanged by the
$e$--$m$ duality. Consequently, only their relative assignment to the plaquette classes is meaningful, while a global interchange $e\leftrightarrow m$ is conventional. This yields the three plaquette assignments summarized in Table~\ref{eq:anyon_regions}, which divide the phase diagram into three regions.

\begin{table}[b]
\centering
\begin{tabular}{|c|c|c|}
\hline
Region & $e$ ($m$) &$m$ ($e$)\\
\hline
$\mathcal A_x$ & $\{\mathrm H,\mathrm S\}$ & $\{\mathrm D\}$\\
$\mathcal A_y$ & $\{\mathrm H,\mathrm D\}$ & $\{\mathrm S\}$ \\
$\mathcal A_z$ & $\{\mathrm S,\mathrm D\}$ & $\{\mathrm H\}$\\
\hline
\end{tabular}
\caption{Sets of plaquettes hosting different types of anyon ($e$ or $m$) in each region $\mathcal A_\alpha$ shown in Fig.~\ref{fig:anyon_regions}. Labels H, S, and D stand for hexagon, square, and dodecagon, respectively. }
\label{eq:anyon_regions}
\end{table}

To determine the boundaries of these regions away from the perturbative regime, we apply the physical-parity criterion of Ref.~\cite{Chern26}.
To summarize, the idea is to consider 
two well-separated vortex configurations that share one fixed reference vortex and a second vortex on two different plaquettes.  Equal physical matter parities identify the two variable vortices with the same species, whereas opposite parities show that they differ by $\psi$ and hence belong to different species. The assignment remains
unchanged under a continuous variation of the couplings as long as the vortex-free fermion gap and the relevant two-vortex gaps remain open.
\begin{figure}[t]
\centering
\includegraphics[width=0.98\columnwidth]{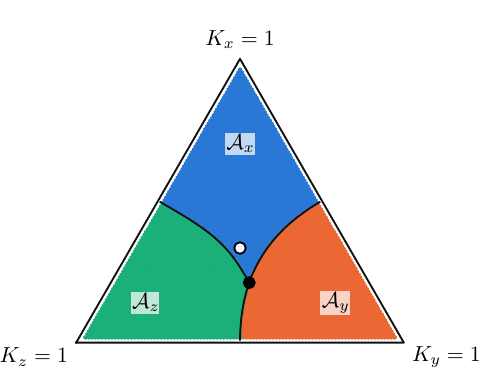}
\caption{Relative vortex-to-anyon assignments in phase diagram, $K_x+K_y+K_z=1$ with
$K_\alpha\geqslant0$. The black curves are boundaries at which the fermion gap closes in the two-vortex sectors containing the
species-changing plaquette class.  They meet at the vortex-free gapless 
point (black dot).  The isotropic point (white dot) lies inside $\mathcal A_x$.}
\label{fig:anyon_regions}
\end{figure}
Each region  $\mathcal A_\alpha$ contains the
corresponding strong-$K_\alpha$ limit, and the
isotropic point lies in $\mathcal A_x$. The black curves in Fig.~\ref{fig:anyon_regions} mark zeros of the fermion gap in the two-vortex sectors whose relative assignment changes.
The vortex-free matter gap is always finite except where the three curves meet at the gapless point.  These boundaries therefore reorganize the microscopic vortex-to-anyon map within one toric code phase.

\section{Kitaev-Heisenberg model}
\label{sec:Kitaev_Heisenberg_SHD}

Having established that the isotropic SHD Kitaev point is a gapped
toric code spin liquid, we now ask how it is destabilized by an isotropic Heisenberg exchange.  The main question is whether the system enters a
conventional ferro- or antiferromagnetic phase directly, or whether the enlarged
SHD unit cell supports intermediate forms of magnetic order.

The Kitaev-Heisenberg model is the minimal setting for this competition. On the honeycomb lattice it produces ferromagnetic, antiferromagnetic,
stripy, and zigzag phases~\cite{Chaloupka10,Chaloupka13,Chaloupka15}.
On the SHD lattice, the coexistence of square, hexagonal, and
dodecagonal plaquettes reorganizes the same exchange competition around
the twelve-site cell.  We use the model as a theoretical probe of this
geometric effect.

\subsection{Hamiltonian and limiting cases}
\label{sec:ham_limits}

The isotropic Kitaev-Heisenberg Hamiltonian is the restriction to 
$K_x=K_y=K_z=K$ of Eq.~\eqref{eq:kh}. We will refer to it as the $(K,J)$ model for simplicity. We parametrize the couplings as
\begin{equation}
(K,J)
=
(\cos\theta,\sin\theta),
\qquad
0\leqslant \theta <2\pi .
\label{eq:parameterization}
\end{equation}
Four special
points provide exact or previously established limits of the phase
diagram.

At $\theta=0$ and $\theta=\pi$, the Heisenberg exchange vanishes and the
model reduces to the isotropic Kitaev model with opposite signs of the
Kitaev coupling.  As shown in Sec.~\ref{sec:Kitaev_SHD}, the isotropic
SHD Kitaev model is gapped and corresponds to a toric code phase.  The
two signs of the isotropic Kitaev coupling have the same free-Majorana spectrum after the sign is absorbed into the
$\mathbb Z_2$ gauge variables, and they represent the two Kitaev limits
around which the spin-liquid regions appear in the phase diagram.  Since
the toric code phase is gapped, it is robust against sufficiently weak local perturbations.  The Heisenberg term in Eq.~\eqref{eq:kh} is therefore
expected to leave finite spin-liquid windows around the Kitaev points, whose phase boundaries are determined numerically in Sec.~\ref{sec:phase_diagram}.

At $\theta=\pi/2$, one has $K=0$ and $J=1$, so that
Eq.~\eqref{eq:kh} becomes the antiferromagnetic Heisenberg model on the
SHD lattice.  This limit is not exactly solvable, but it has been
studied previously using exact diagonalization, coupled-cluster
methods, and series expansions~\cite{Tomczak01,Richter04,Farnell14,Farnell18}.
This will serve as a limit benchmark for the tensor-network and ED calculations in the
antiferromagnetically ordered regime that we present in Appendices \ref{app:gpeps} and \ref{app:ed}.

At $\theta=3\pi/2$, one has $K=0$ and $J=-1$, and the model is the ferromagnetic Heisenberg model, whose ground-state manifold consists of all states with maximal total spin with energy $E_{\rm FM_+}=-3N/8$, where $N$ is the total number of lattice sites.

\subsection{Order parameters and classical phases}
\label{sec:classical_phase_diagram}

We first consider the classical ground state of the Kitaev-Heisenberg Hamiltonian (\ref{eq:kh}). We will design collinear classical spin ansatze and minimize the energy in this ansatze space to find the optimal collinear state. Then, we will show that their energy saturates the lower bound, proving that these collinear states are the global classical ground state of the model.

\begin{figure}[t]
\centering
\includegraphics[width=\columnwidth]{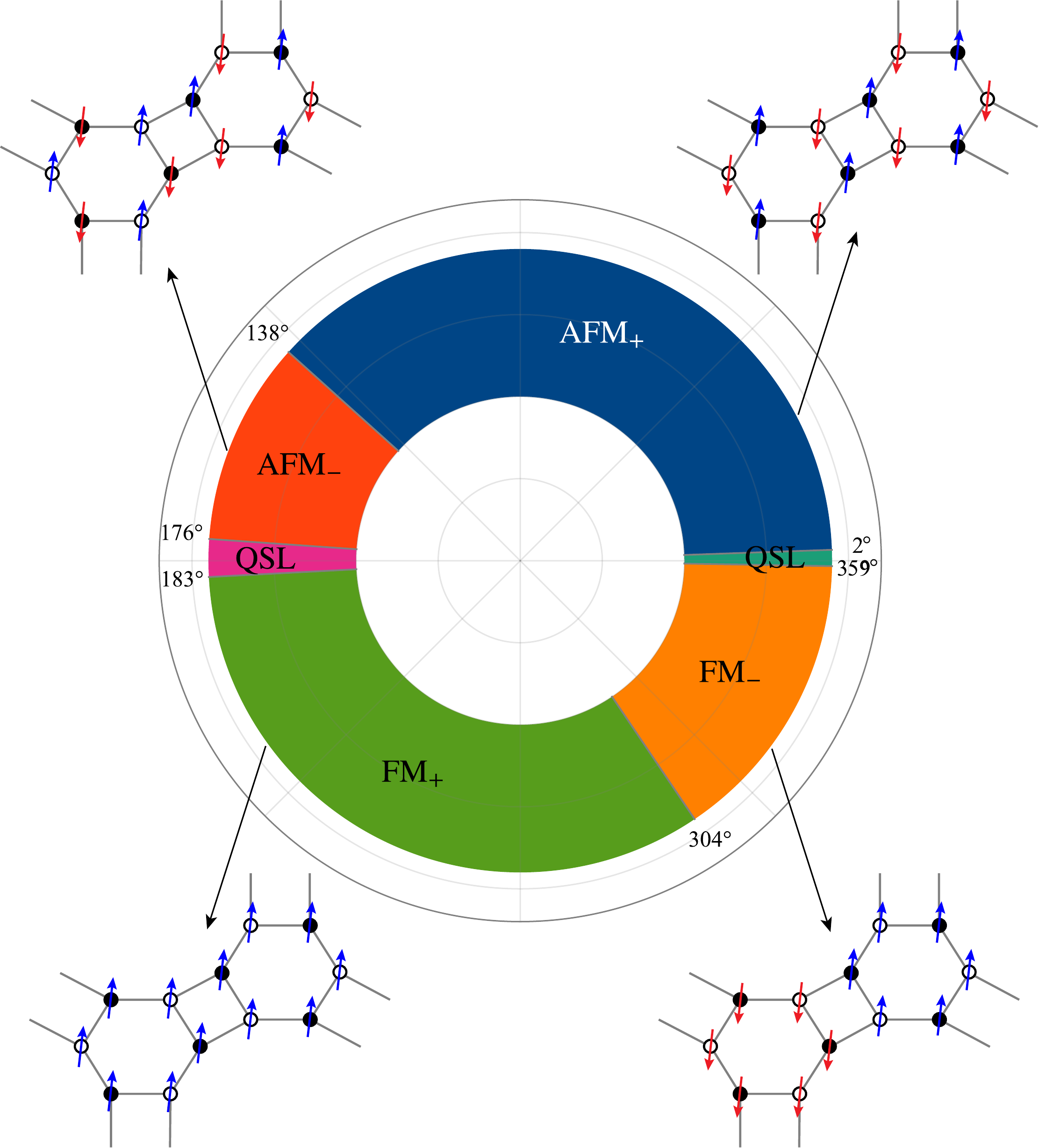}
\caption{Ground-state phase diagram of the isotropic
Kitaev-Heisenberg model on the SHD lattice,
$(K,J)=(\cos\theta,\sin\theta)$, obtained from gPEPS.  The two narrow QSL phases around $\theta=0,\pi$ correspond to toric code topological phases, as inferred from the exactly solvable pure Kitaev limit \mbox{$(J=0)$}. The remaining regions are the four magnetic phases, FM$_\pm$ and AFM$_\pm$, discussed in Sec.~\ref{sec:classical_phase_diagram}.  
}
\label{fig:phase_diagram}
\end{figure}

We consider collinear classical spin configurations of the form
\begin{equation}
\mathbf S_i
=
S f_i^\mu \hat{\mathbf n},
\label{eq:classical_collinear_ansatz}
\end{equation}
where $S$ is the classical spin length, $\hat{\mathbf n}$ is a unit vector in spin space specifying the collinear spin orientation, and the sign $f_i^\mu=\pm1$ taken from one of the following four ansatze: 
\begin{subequations}
\label{eq:classical_sign_patterns}
\begin{align}
f_i^{{\rm AFM}_+} &= \zeta_i,
\label{eq:f_AFMp}
\\
f_i^{{\rm AFM}_-} &= \eta_i\zeta_i,
\label{eq:f_AFMm}
\\
f_i^{{\rm FM}_+} &= 1,
\label{eq:f_FMp}
\\
f_i^{{\rm FM}_-} &= \eta_i.
\label{eq:f_FMm}
\end{align}
\end{subequations}
Here $\eta_i$ is a hexagon index with $\eta_i=\left\{\begin{smallmatrix} +1 & i \in A \text{ hexagon} \\ -1 & i \in B \text{ hexagon}\end{smallmatrix}\right.$  [see
Fig.~\ref{fig:ed_36_cluster}(a)], and $\zeta_i$ is a sublattice variable defined as $\zeta_i = \left\{\begin{smallmatrix} +1 & i \in \text{ white sites} \\ -1 & i \in \text{ black sites}\end{smallmatrix}\right.$ (see Fig.~\ref{fig:kekule_coloring}).  With this convention, the labels FM and AFM refer to the local arrangement on a hexagon, while the
subscripts $+$ or $-$ distinguish whether this hexagonal pattern is arranged uniformly or staggered between neighboring hexagons (see Fig.~\ref{fig:phase_diagram} for illustration).
Thus, AFM$_+$ is the usual bipartite N\'eel pattern, and FM$_+$ is the
usual ferromagnet.  The two ``minus'' phases differ from these
conventional orders by reversing the relative sign on the $z$ bonds, which connect the two hexagons in the twelve-site unit cell.


Minimizing the energy over the four sign patterns $\mu$ in
Eq.~\eqref{eq:classical_sign_patterns} and over the unit vector
$\hat{\mathbf n}$, we obtain
\begin{subequations}
\label{eq:gse_cl}
\begin{align}
E_{{\rm AFM}_+}^{\rm cl}
&=
NS^2
\left(
-\frac{K}{2}
-
\frac{3J}{2}
\right),
\label{eq:gse_AFMp_cl}
\\
E_{{\rm AFM}_-}^{\rm cl}
&=
NS^2
\left(
-\frac{|K|}{2}
-
\frac{J}{2}
\right),
\label{eq:gse_AFMm_cl}
\\
E_{{\rm FM}_+}^{\rm cl}
&=
NS^2
\left(
\frac{K}{2}
+
\frac{3J}{2}
\right),
\label{eq:gse_FMp_cl}
\\
E_{{\rm FM}_-}^{\rm cl}
&=
NS^2
\left(
-\frac{|K|}{2}
+
\frac{J}{2}
\right),
\label{eq:gse_FMm_cl}
\end{align}
\end{subequations}
where $N$ is the total number of lattice sites.

These energy branches from the collinear ansatze in fact give rise to \emph{global} classical ground state. To establish that these states are global classical ground states, we
bound the classical energy without restricting the spin configurations to be collinear.  Applying the inequality \mbox{$cab\geqslant -|c|(a^2+b^2)/2$} to each component on every bond yields a lower bound on the classical energy
\begin{equation}
\begin{aligned}
&\sum_{\alpha=x,y,z}\sum_{\langle ij\rangle_\alpha}
\sum_{\rho=x,y,z}
(J+K\delta_{\alpha\rho})S_i^\rho S_j^\rho
\\
&\geqslant
-\frac{1}{2}
\sum_{\alpha}\sum_{\langle ij\rangle_\alpha}
\sum_{\rho=x,y,z}
|J+K\delta_{\alpha\rho}|
\bigl[(S_i^\rho)^2+(S_j^\rho)^2\bigr]\\
&=-\frac{1}{2}
\sum_i\sum_{\rho=x,y,z}
\left(\sum_{\alpha=x,y,z}|J+K\delta_{\alpha\rho}|\right)
(S_i^\rho)^2
\\
&=
-\frac{NS^2}{2}\bigl(|J+K|+2|J|\bigr).
\end{aligned}
\label{eq:classical_global_bound}
\end{equation}
Substituting the four patterns in Eq.~\eqref{eq:classical_sign_patterns} with the spin directions in Table~\ref{tab:classical_gs} shows that each of them saturates this bound in its stability region.  Their energies in Eq.~\eqref{eq:gse_cl} are therefore the global classical ground-state energies.

The phase sequence exposes two distinct levels of magnetic
organization.  The AFM/FM label records whether the six spins within a
hexagon form a local N\'eel or ferromagnetic moment, while the subscript
records the relative phase of that moment on neighboring hexagons.  The
AFM$_+$ /AFM$_-$ transition therefore reorganizes the alignment of
pre-existing hexagonal N\'eel units without removing their local N\'eel
character; the same statement applies to the FM$_+$/ FM$_-$ transition.  The
intermediate "$-$" phases are consequently tied to the hexagonal
building blocks of the SHD lattice and provide a geometric alternative
to the stripy and zigzag orders observed in the honeycomb model~\cite{Chaloupka13}.

\begin{table}[b]
\centering
\caption{Classical ground states of the SHD
Kitaev-Heisenberg model.}
\label{tab:classical_gs}
\begin{tabular}{|c|c|c|c|}
\hline
Order & Collinear direction $\hat{\mathbf{n}}$ & Stability region\\
\hline
AFM$_+$  & arbitrary & $0^\circ<\theta<135^\circ$\\
AFM$_-$  & $\hat{\mathbf n}=\hat{\mathbf z}$ & $135^\circ<\theta<180^\circ$\\
FM$_+$  & arbitrary & $180^\circ<\theta<315^\circ$\\
FM$_-$  & $\hat{\mathbf n}=\hat{\mathbf z}$ & $315^\circ<\theta<360^\circ$\\
\hline
\end{tabular}
\end{table}


\subsection{Quantum phase diagram}
\label{sec:phase_diagram}

Having established the candidate classical orders and their
corresponding diagnostics, we now determine the quantum phase diagram
along $(K,J)=(\cos\theta,\sin\theta)$.  As was the classical case, the competing SHD orders are
distinguished by both the spin pattern within each hexagon and the
relative alignment of neighboring hexagons.

To compute the phase diagram, we employ a variational tensor-network method using graph-based projected entanglement-pair states (gPEPS) with a bond dimension of $D=8$. The gPEPS results are benchmarked against exact diagonalization (ED) on a $N=36$ spins system with periodic boundary conditions. Numerical details are given in
Appendices~\ref{app:gpeps} and \ref{app:ed}. 
%
\begin{figure}[t]
\centering
\includegraphics[width=\columnwidth]{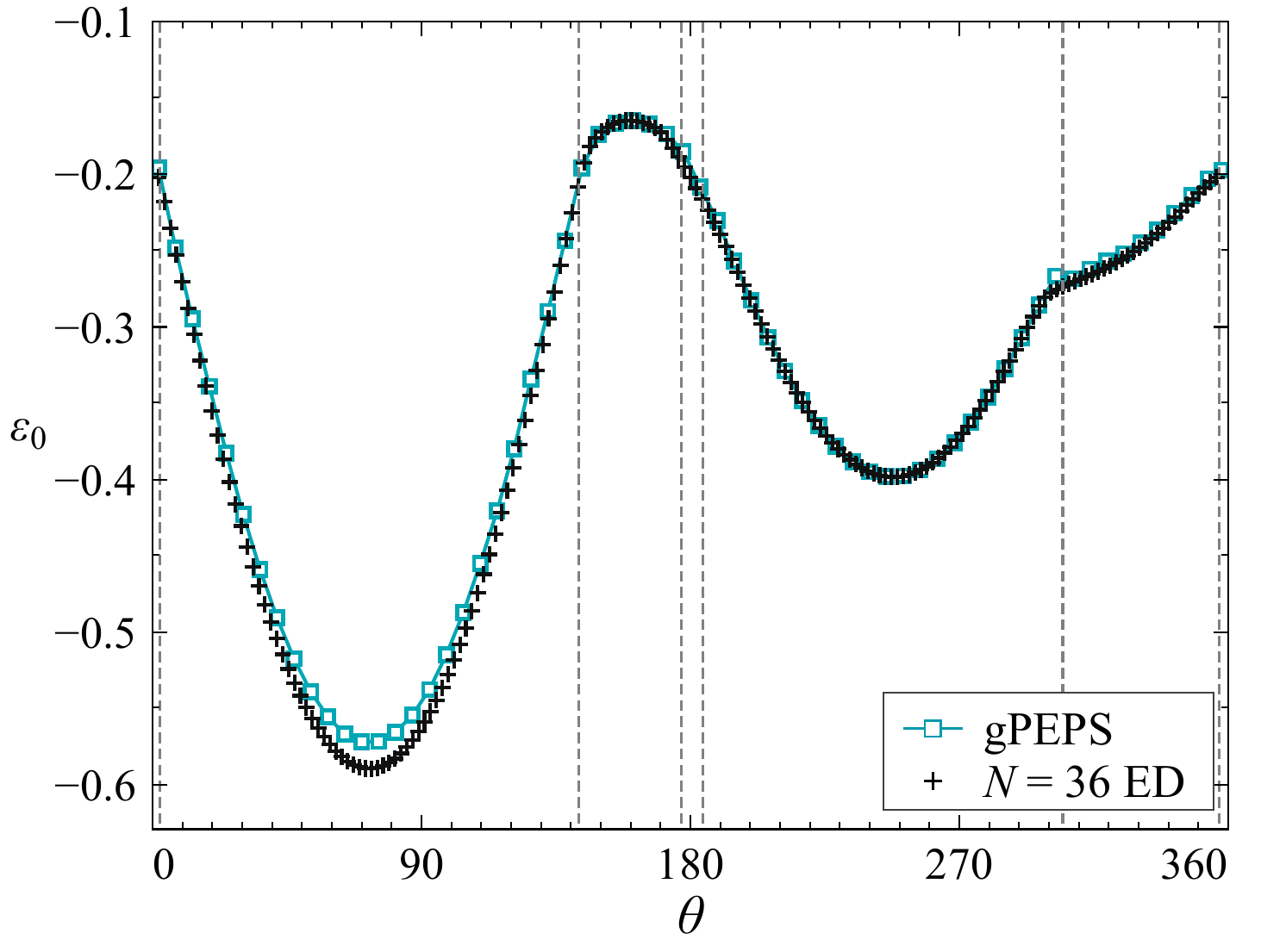}
\caption{Ground-state energy per site  $\varepsilon_0$ of the isotropic Kitaev-Heisenberg model on the SHD lattice.  The gPEPS data are compared with ED ($N=36$) data. Vertical dashed lines mark the approximate phase boundaries inferred
from the behavior of the order parameters shown in Fig.~\ref{fig:order_params}. The black star marks the exact product-state benchmark at $\theta_\star=\pi-\tan^{-1}(1/2)$, where $\varepsilon_0=-3/(8\sqrt5)$.}
\label{fig:energy_kh}
\end{figure}
%

We first present in Fig.~\ref{fig:energy_kh} the ground-state energies obtained from gPEPS and ED.
We benchmark the gPEPS and ED at the Heisenberg points $(K,J) = (0,\pm 1)$. At the ferromagnetic point $J = -1$, both gPEPS and ED reproduce the exact
ground-state energy per site $\varepsilon_0=-3/8$. At the antiferromagnetic point $J=1$, the gPEPS calculations give $\varepsilon_0 = -0.553685$, which is in agreement with previous numerical studies~\cite{Tomczak01,Richter04,Farnell14,Farnell18}. For the $N= 36$ cluster shown in Fig.~\ref{fig:ed_36_cluster}(b), the ED result is $\varepsilon_0=-0.559677$~\cite{Tomczak01}. Furthermore, for all values of $\theta$, the two energy curves track each other closely, showing the validity of the infinite-system gPEPS and the finite-size ED.

To identify the different phases, we turn to the study of quantum order parameters. We define
\begin{equation}\label{eq:quantum_op}
O_\mu = \left| \frac{1}{N}
\sum_i f_i^\mu \langle \mathbf{S}_i\rangle \right|,
\end{equation}
where $\mu = \{\mathrm{FM}_+,\mathrm{FM}_-,\mathrm{AFM}_+,\mathrm{AFM}_-\}$ and $f_i^\mu$ is defined in Eq.~\eqref{eq:classical_sign_patterns}:
namely, the AFM/FM label selects the local hexagon moment, and the $+/-$
subscript selects uniform/staggered inter-hexagon alignment.  These are
the thermodynamic counterparts of the classical sign patterns in
Sec.~\ref{sec:classical_phase_diagram}. In the infinite-system  gPEPS
calculation, the optimized state can select a symmetry-broken
representative, so the onsite moments $\langle \mathbf S_i\rangle$ are
directly accessible.  Finite-size ED instead uses the corresponding
root-mean-square correlations, as described in Appendix~\ref{app:ed}.

\begin{figure}[t]
\centering
\includegraphics[width=\columnwidth]{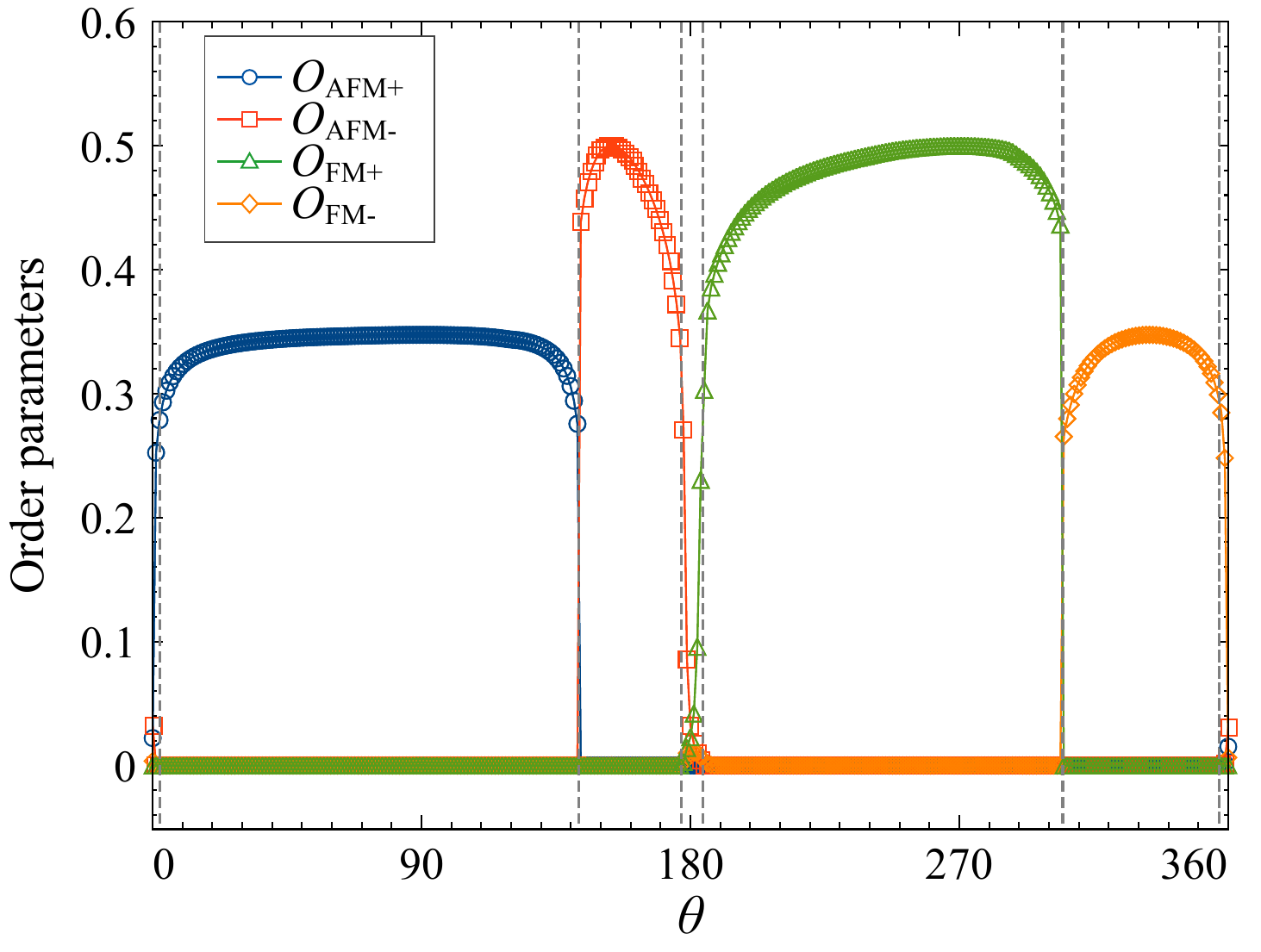}
\caption{Hexagon-resolved magnetic order parameters obtained from gPEPS.
The quantities are defined in Eq.~\eqref{eq:quantum_op}; vertical dashed lines mark the phase boundaries used in
Fig.~\ref{fig:phase_diagram}.}
\label{fig:order_params}
\end{figure}

The resulting order parameters from gPEPS are shown in Fig.~\ref{fig:order_params}; those from ED are given in the Appendix~\ref{app:ed}. Both methods confirm that the phase diagram contains six
regions: the two regions
adjacent to the Kitaev limits are quantum spin liquids continuously
connected to the exactly solved toric code points of
Sec.~\ref{sec:Kitaev_SHD}; the
remaining four phases are magnetically ordered and are the finite $S$ counterpart of the classical phases discussed in Sec.~\ref{sec:classical_phase_diagram}, hence we address them using the same classical label $\mathrm{FM}_{\pm}$ and $\mathrm{AFM}_{\pm}$. The conventional
AFM$_+$ and FM$_+$ phases occur around the antiferromagnetic and
ferromagnetic Heisenberg limits, respectively, while AFM$_-$ and FM$_-$
appear as intermediate phases between the conventional orders and the Kitaev spin liquids. These phases are illustrated in Fig.~\ref{fig:phase_diagram}.

In addition to gPEPS and ED, we perform a semiclassical linear-spin wave theory (LSWT) analysis to estimate the quantum fluctuations due to the finite spin $S=1/2$. We find that the energy calculated from LSWT agrees well with the ground-state energy computed from gPEPS and the $N = 36$ ED deep in the magnetic phases, indicating that quantum fluctuations are negligible inside these phases; whereas at the phase boundary, the LSWT reveals large quantum fluctuations. The detailed calculations can be found in Appendix \ref{app:lswt}.

Narrow spin-liquid regions appear around $\theta=0$ and $\pi$.  The
order parameters in Fig.~\ref{fig:order_params} vanish in these windows,
and the optimized tensors retain the local structure expected near the
Kitaev limit: as an example, near $\theta = \pi$, on an $\alpha$-type bond the dominant nearest-neighbor
correlator is the $\alpha$ spin component.  We identify these regions as
the continuation of the exactly solvable toric code phase established in Sec.~\ref{sec:Kitaev_SHD}. 

\subsection{Exact ground state at $(K,J)=(-2,1)$}
\label{sec:exact_product_state}

A particular point inside the AFM$_-$ region, \mbox{$(K,J)=(-2,1)$}, is interesting: the ground state can be exactly solved, hence this parameter point provides an exact reference for the AFM$_-$ phase.  In the normalized parametrization of Eq.~\eqref{eq:parameterization}, this coupling ratio corresponds to
$\theta_\star=\pi-\tan^{-1}(1/2)\simeq153.4^\circ$, where the exact energy per site is
$\varepsilon_0=-3/(8\sqrt{5})\simeq-0.167705$. To see this, we pick a bond $\langle ij\rangle_\alpha$ and let
$\{\alpha,\beta,\gamma\}=\{x,y,z\}$.  The corresponding term in the Hamiltonian defines a rank-one projector 
\begin{equation}
\begin{aligned}
h_{ij}^{(\alpha)}
&=\mathbf S_i\cdot\mathbf S_j-2S_i^\alpha S_j^\alpha
=P_{ij}^{(\alpha)}-\frac14,\\
P_{ij}^{(\alpha)}
&=
\frac{1}{4}
\left(
\mathds{1}
-
\sigma_i^\alpha\sigma_j^\alpha
+
\sigma_i^\beta\sigma_j^\beta
+
\sigma_i^\gamma\sigma_j^\gamma
\right),
\end{aligned}
\label{eq:exact_projector}
\end{equation}
giving the following lower bound for the Hamiltonian at $(K,J) = (-2,1)$
\begin{equation}
H = -\frac{3N}{8}
+ \sum_{\alpha=x,y,z}\sum_{\langle ij\rangle_\alpha}
P_{ij}^{(\alpha)}\geqslant -\frac{3N}{8}.
\label{eq:frustration_free}
\end{equation}
Saturating this inequality requires a common zero eigenvalue of all bond projectors.

We now construct such a state using the AFM$_-$ sign pattern introduced in Sec.~\ref{sec:classical_phase_diagram}.  Let
$f_i^{{\rm AFM}_-}=\pm 1$ denote the AFM$_-$ pattern, and define
the spin-coherent product state
\begin{equation}
\ket{\Psi_{{\rm AFM}_-}}
=
\bigotimes_i
\ket{f_i^{{\rm AFM}_-}\hat{\mathbf z}}_i.
\label{eq:exact_AFMm_state}
\end{equation}
For each bond $\langle i j\rangle_\alpha$, one has

\begin{equation}
f_i^{{\rm AFM}_-}f_j^{{\rm AFM}_-}
=
\begin{cases}
-1, & \alpha=x,y,\\
+1, & \alpha=z,
\end{cases}
\end{equation}
and one can verify that
$P_{ij}^{(\alpha)}|\Psi_{{\rm AFM}_-}\rangle
=0$ for every bond $\langle ij\rangle_\alpha$,
so that $|\Psi_{{\rm AFM}_-}\rangle$ is an exact ground state.


This construction gives one exact ground state, and a global spin flip via time-reversal symmetry gives a second. Appendix~\ref{app:exact_solvable_point} shows that these two states exhaust the ground-state manifold and identifies a hidden $U(1)$ symmetry at this point.
-


\subsection{Absence of Klein duality}

Finally, unlike the original Kitaev--Heisenberg model on the honeycomb lattice~\cite{Chaloupka10,Chaloupka13,Chaloupka15}, the SHD phase diagram is not organized by a global Klein duality. A necessary condition for such a duality is that the parities of the three bond types be identical on every plaquette~\cite{Kimchi14}. This condition fails for the present coloring because each hexagon contains three $x$ bonds, three $y$ bonds, and no $z$ bonds [Fig.~\ref{fig:kekule_coloring}]. Nevertheless, at the isolated point $K=-2J$, a Klein-type sublattice rotation survives and exposes the hidden $U(1)$ symmetry derived in Appendix~\ref{app:exact_solvable_point}. On the conventionally colored honeycomb lattice, the same coupling ratio is Klein-dual to the ferromagnetic Heisenberg point and therefore has hidden $SU(2)$ symmetry~\cite{Chaloupka15}.

\section{Summary and outlook}
\label{sec:conclude}

In this article, we studied the Kitaev–Heisenberg model for spin-$1/2$ systems on the SHD lattice, which is one of the four tricoordinated Archimedean lattices. We first analyzed the pure Kitaev model ($J=0$) for a specific three-edge coloring, shown in Fig.~\ref{fig:kekule_coloring}, and computed its ground-state phase diagram. This phase diagram consists of a unique toric code phase with an isolated gapless point (see Fig.~\ref{fig:Kitaev_PD}). At this point, the Majorana band spectrum exhibits a single Dirac point, as shown in Fig.~\ref{fig:majorana_gapless_dispersion}. Following the recent work by Chern et al.~\cite{Chern26}, we also provided the toric code anyon assignment for each plaquette as a function of the coupling parameters (see Fig.~\ref{fig:anyon_regions} and Table~\ref{eq:anyon_regions}). In a second step, we considered the isotropic Kitaev model, which is gapped in the present case, and introduced a competing Heisenberg term. Combining several complementary methods (gPEPS, ED, and LSWT), we computed the phase diagram shown in Fig.~\ref{fig:phase_diagram}. In addition to the QSL phases originating from the Kitaev limits, this diagram contains four distinct magnetically ordered phases, which we clearly identified by relating them to their classical counterparts. 

The properties of the Kitaev model depend on two important parameters: the geometry of the lattice and the three-edge coloring, each of which can lead to very different phases. An interesting open question is to determine, for a given lattice geometry, the “optimal coloring,” i.e., the one that yields the lowest ground-state energy. This question was recently addressed in Refs.~\cite{Shixiong_thesis} for the honeycomb lattice, for which the Kekul\'e coloring~\cite{Kamfor10} was found to be the best candidate for any values of the couplings $K_\alpha$. Conversely, in the SHD lattice, the optimal coloring seems to depend on the $K_\alpha$'s~\cite{Shixiong_thesis}. Determining the optimal coloring over the entire parameter range is an interesting open problem. Along the same lines, one may ask about the optimal geometry, for instance, at the isotropic point, where the coloring becomes irrelevant. In other words, which tricoordinated lattice has the lowest ground-state energy?

Possible realizations of the Kitaev--Heisenberg model on the SHD lattice would
certainly be challenging.  To our knowledge, spin-$1/2$ magnets on the SHD
lattice have not yet been identified. Nevertheless, the SHD lattice structure
itself has been realized chemically, e.g., in certain metal--organic
frameworks~\cite{Eddaoudi02}, or as patterned structures using Co/Pt
nanomagnets~\cite{Pac26}.  The demanding task would be to stabilize the SHD
lattice with magnetic ions and sufficiently strong spin--orbit coupling to
generate bond-dependent exchange of Kitaev type.

Another interesting direction is the deformation of the Kitaev model and the
possible resulting phases.  In the original Kitaev honeycomb
model, further-neighbor couplings and a magnetic field produce a large set of
phases in the sixteenfold way~\cite{Kitaev06,Zhang20,Fuchs20}.  A similar study
has been carried out for the Kekul\'e model~\cite{Quinn15,Mirmojarabian20}. In
the Kitaev SHD model, the isotropic point is already gapped with $\nu=0$ and
hence remains in the same phase under a sufficiently weak field, whereas our
preliminary calculation at the isolated gapless point shows that a field along
the $[111]$ direction opens a topologically trivial gap with $\nu=0$. It is
therefore interesting to ask whether a non-Abelian chiral phase with a nonzero
$\nu$ could be produced by additional perturbations.

Finally, we focused on the phases of the Kitaev--Heisenberg model in this work,
but have not probed the nature of the phase transitions.
Figure~\ref{fig:phase_diagram} exhibits four spin-liquid--to--magnetic phase
boundaries.  A standard physical picture asserts that these transitions are
driven by condensation of the lowest-energy bosonic anyon, or of a spin-carrying
flux--Majorana bound state descended from it.  Whether these transitions are
continuous or first order is an interesting question for future studies.
Useful precedents are the three-dimensional XY$\times\mathbb Z_2$ transition proposed for
the perturbed Kekul\'e--Kitaev model~\cite{Quinn15}, the self-dual Abelian Higgs
description of a Kitaev spin liquid condensing into a magnet~\cite{Nanda20},
and the XY$^*$ transitions generated by symmetry-fractionalized anyon
condensation~\cite{Schuler23}.  For the SHD model, detailed numerical studies
of the order parameters, correlation lengths, excitation gaps, and
entanglement across each boundary will be needed to distinguish between these scenarios.

\section*{Acknowledgments}

The work of Y.I. was performed, in part, at the Aspen Center for Physics, which is supported by a grant from the Simons Foundation (1161654, Troyer). This research was also supported in part by grant NSF PHY-2309135 to the Kavli Institute for Theoretical Physics. Y.I. acknowledges support from the Abdus Salam International Centre for Theoretical Physics through the Associates Programme, from the Simons Foundation through Grant No.~284558FY19, from IIT Madras through the Institute of Eminence program for establishing QuCenDiEM (Project No. SP22231244CPETWOQCDHOC), and the International Centre for Theoretical Sciences for participation in the Discussion Meeting --- Fractionalized Quantum Matter (code: ICTS/DMFQM2025/07). S.S.J. and J.V. acknowledge support from IIT Madras as Visiting Faculty Fellows during which this project was initiated.

\section*{Data Availability Statement}
The data and analysis scripts used to generate the results are available on Zenodo public repository at
\href{https://doi.org/10.5281/zenodo.21261528}{10.5281/zenodo.21261528}. Additional data
are available from the corresponding author upon reasonable request.

\appendix
\section{Graph-based PEPS method}
\label{app:gpeps}

We obtain the thermodynamic-limit energy and magnetic order parameters
with a graph-based projected entangled-pair state (gPEPS) on the
infinite SHD lattice.  The ansatz uses the twelve-site unit cell
and its tricoordinated connectivity directly.  The bond dimension $D$
controls the variational manifold, while the environment dimension
$\chi$ controls the accuracy of the final contraction.  ED and LSWT
provide the finite-size and semiclassical benchmarks described in
Appendices~\ref{app:ed} and \ref{app:lswt}.

\subsection{Tensor-network ansatz}
We use an infinite PEPS ~\cite{Verstraete04,Verstraete08,Orus14} with a twelve-site unit cell matching the primitive cell of the SHD lattice in Fig.~\ref{fig:kekule_coloring}. Since the lattice is tricoordinated, each site tensor has one physical spin-$1/2$ index and three virtual indices:
\begin{equation}
(T_a^{s})_{l_1 l_2 l_3},
\qquad
s=1,2,
\qquad
l_m=1,\ldots,D,
\label{eq:gpeps_tensor}
\end{equation}
where $s$ is the physical dimension and $D$ is the virtual bond dimension. We denote the twelve tensors by $T_{A_0},\ldots,T_{A_5},T_{B_0},\ldots,T_{B_5}$, following the labeling in Fig.~\ref{fig:ed_36_cluster}(a), where $A$ and $B$ identify the two hexagons within the unit cell. A structure matrix specifies the connections between virtual legs and the $x$, $y$, or $z$ color of each bond~\cite{Jahromi19,Jahromi20,Jahromi21}. This construction directly represents the connectivity of the SHD lattice without embedding the model in an auxiliary square lattice for the simple update of tensors.

\subsection{Imaginary-time optimization}
We optimize the tensors by imaginary-time evolution using the simple-update scheme~\cite{Jordan08,Jahromi19}. Starting from a random initial tensor network, we approximate the ground state through the projection
\begin{equation}
|\Psi_0\rangle
\propto
\lim_{\tau\rightarrow\infty}
{\rm e}^{-\tau H}
|\Psi_{\rm init}\rangle,
\label{eq}
\end{equation}
which requires a nonzero overlap between the initial state and the ground-state subspace. In practice, the evolution is performed starting from a PEPS with bond dimension $D=2$ that grows during the update to $D_{\rm max}$. We approximate the evolution operator using a second-order Suzuki--Trotter decomposition~\cite{Suzuki76}, with each sweep covering the update of all eighteen inequivalent bonds in the twelve-site unit cell.

\begin{figure}[t]
\centering
\includegraphics[width=\columnwidth]{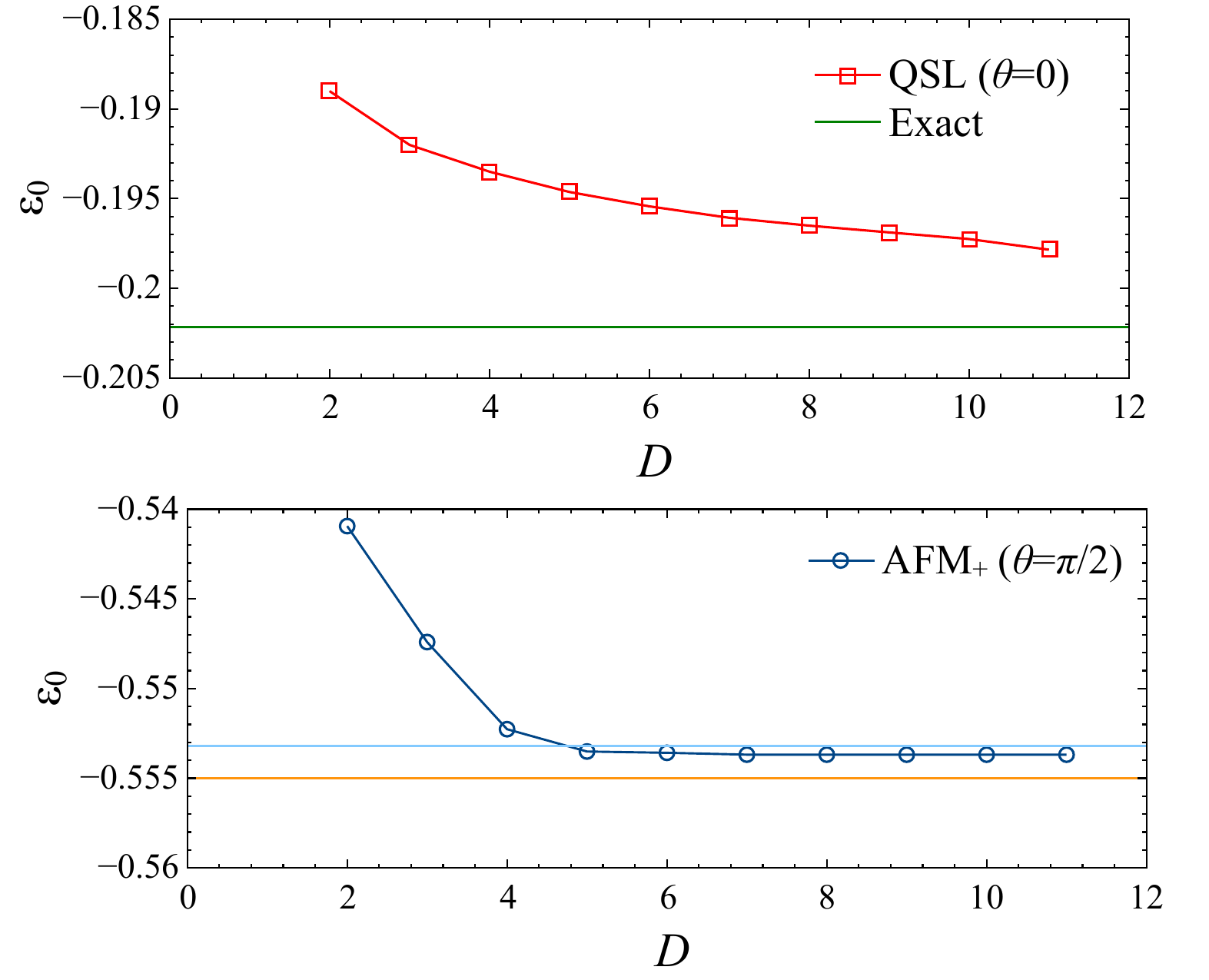}
\caption{Bond-dimension dependence of the gPEPS ground-state energy per
site at two benchmark points.  (a) Energy at the isotropic Kitaev point
$\theta=0$, compared with Eq.~\eqref{eq:gse_exact}.  (b) Energy at the
antiferromagnetic Heisenberg point \mbox{$\theta=\pi/2$}, compared with
previous numerical estimates for the SHD Heisenberg antiferromagnet
from Refs.~\cite{Tomczak01,Farnell18}.}
\label{fig:energy_isotropic}
\end{figure}

Starting from a time step $\delta\tau=10^{-1}$, we progressively reduce it down to $10^{-3}$, performing up to $3000$ sweeps at each time step. Convergence is monitored through the energy per site and the local observables entering the order parameters. The largest bond dimension in this study is $D_{\rm max}=8$ for the high-resolution data in Fig.~\ref{fig:energy_kh}, \ref{fig:order_params} and $D_{\rm max}=11$ for the energy scalings in Fig.~\ref{fig:energy_isotropic}.

\subsection{Contraction and observables}

We evaluate expectation values by contracting the infinite double-layer tensor network formed from the optimized ket and bra tensors using the corner transfer matrix renormalization group (CTMRG) method~\cite{Nishino96,Orus09} by adopting the network into brick-wall lattice with dummy links to form a square geometry (see Ref.~\cite{jahromi2018} for details). The accuracy of the CTMRG algorithm is controlled by the
environment dimension $\chi$. In the calculations reported in this work, we used $\chi=D^2$ for $D\leqslant8$ and then set it to $\chi=64$ for larger $D$ due to computational resource constraints. 

Once the converged environment is obtained, the one-site reduced density matrix gives the local magnetization $\mathbf m_a=\langle \mathbf S_a\rangle$ for each of the twelve inequivalent sites in the unit cell.  These
onsite moments are then entered into the order parameters in
Eqs.~\eqref{eq:quantum_op} to characterize the underlying phases in the phase diagram.

In the spin-liquid windows all four order parameters are suppressed, and the dominant correlation on a  bond $\alpha$ remains the spin component $\alpha$.  This provides a local tensor-network
diagnostic of the proximity to the exactly solvable Kitaev limit, although the gPEPS calculation by itself is not used to establish topological order.

\subsection{Benchmark calculations}

Fig.~\ref{fig:energy_isotropic} presents two benchmarks of the gPEPS calculation. At the isotropic Kitaev point, $\theta=0$, the energy per site approaches the exact value in Eq.~\eqref{eq} as $D$ increases, supporting the accuracy of the tensor-network ansatz.

At $\theta=\pi/2$, the model reduces to the antiferromagnetic Heisenberg model on the SHD lattice. The energy converges toward previously reported numerical values~\cite{Tomczak01,Farnell18}, while the local moments exhibit the AFM$_+$ pattern. This provides a complementary benchmark of the optimization in a conventionally ordered regime.

At each sampled coupling, we identify the magnetic ordering pattern from the dominant pattern-resolved order parameter. All four order parameters vanish within numerical resolution in the spin-liquid windows.

\section{Exact diagonalization}
\label{app:ed}

ED provides a finite-size system benchmark and pattern-resolved spin
correlations for the phase diagram in Sec.~\ref{sec:phase_diagram}.  A
symmetry-preserving eigenstate has vanishing onsite moments, so
magnetic tendencies are measured through invariant two-point functions.

\begin{figure}[t]
\centering
\includegraphics[width=0.98\columnwidth]{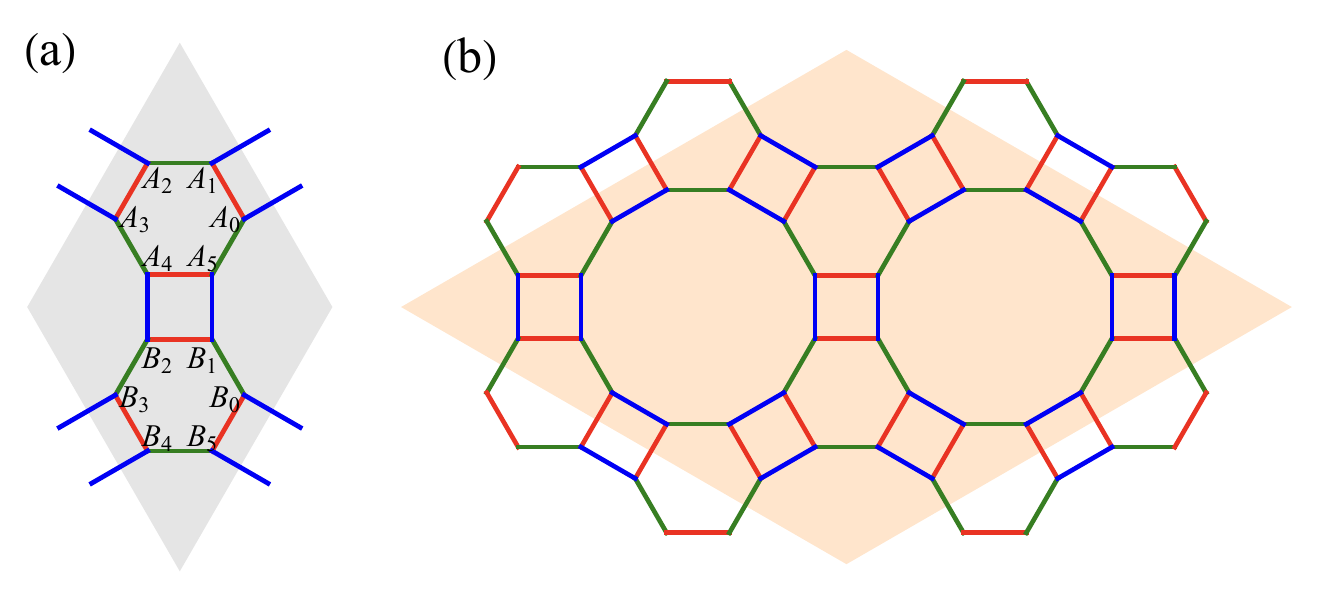}
\caption{(a) Labeling of the twelve sites in the unit cell used in the Kitaev and spin-wave calculations; see Eq.~\eqref{eq:appC_basis_order}. The unit cell is shown in gray. (b) Geometry of the $N=36$ SHD cluster used for exact
diagonalization.  The cluster is placed on a torus that contains three twelve-site unit cells (the region shown in orange).  Colored bonds indicate the three Kitaev bond types of the coloring used throughout the paper, and periodic boundary conditions identify sites and bonds across opposite
edges of the torus.}
\label{fig:ed_36_cluster}
\end{figure}

We use the $N=36$ cluster shown in Fig.~\ref{fig:ed_36_cluster}(b).  The
cluster contains three twelve-site unit cells of the SHD lattice and is
equipped with periodic boundary conditions.  
The full Hilbert-space dimension is $2^{36}$, which is too large for a direct diagonalization.  We therefore
exploit the exact symmetries of the cluster to block-diagonalize
the Hamiltonian.  The symmetry group used in the
ED calculation is
\begin{equation}
G_{\rm cluster}
=
\left(\mathbb Z_3 \rtimes D_6\right)
\times
\mathbb Z_2^2,
\label{eq:ed_cluster_symmetry}
\end{equation}
where the first factor $\mathbb Z_3$ is the translation group of the finite cluster (which contains three unit cells), $D_6$ is the point group generated by $C_3$, $C_2$, and $M_x$, and $\mathbb{Z}_2^2$ are the global spin rotation group (see 
Sec.~\ref{sec:projective_symmetry} for their definition).

Our $N= 36$ ED uses the trivial irreducible representation (irrep) of the group $G_{\rm cluster}$.  This is
the sector expected for the symmetric finite-size combination of
symmetry-related ordered states. The validity of searching for ground states in this sector is benchmarked by the following three complementary checks. First, on the $N=24$ torus we
diagonalized every symmetry sector throughout the angular scan and
selected the global minimum.  
Second, on the $N=36$ torus we performed a
complete irrep comparison at the representative FM$_-$ coupling $(K,J)=(2,-1)$ and found the ground state in the trivial irrep sector.  
Third, at the antiferromagnetic Heisenberg point our
trivial-sector value $E_0/N=-0.559677$ agrees with the published
full-ED result for the same cluster~\cite{Tomczak01}.  Additional
nontrivial $N=36$ sectors were checked at other representative couplings.

Although $\langle\mathbf S_i\rangle=0$ in the fully symmetric
eigenstate sector, magnetic correlations remain finite. This is exactly how we built the ED parameter: it is the root mean square of a correlation function
\be
\left(O_{\mu}^{\rm ED}\right)^2
=
\frac{1}{N^2}
\sum_{i,j}
f_i^\mu f_j^\mu
\left\langle
\mathbf S_i\cdot \mathbf S_j
\right\rangle.
\label{eq:ed_rms_order_parameter}
\ee
This is also recognized as the structure factor for $f_i^\mu$. At finite $N$, the onsite term
$\langle\mathbf S_i^2\rangle=S(S+1)$ gives
$O_\mu^{\rm ED}=\sqrt{S^2+S/N}$ for a maximally polarized product state,
so a root mean square can lie slightly above $S$ before approaching it as
$N\to\infty$.  This two-point quantity is the ED counterpart of the
onsite-moment order parameter used in gPEPS.

\begin{figure}[t]
\centering
\includegraphics[width=\columnwidth]{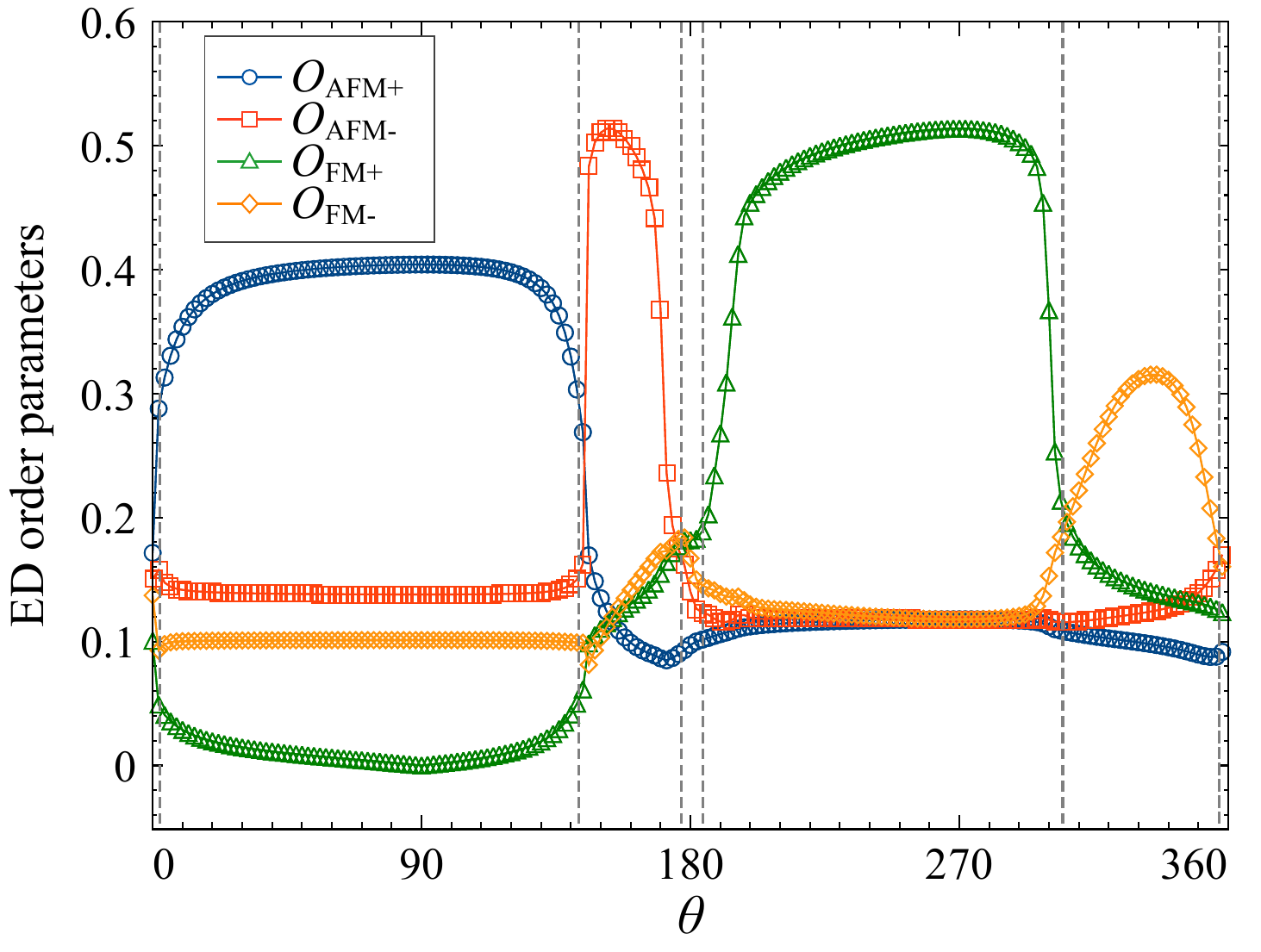}
\caption{Root-mean-square order parameters obtained from the
$N=36$ exact-diagonalization calculation in the trivial symmetry sector.
The four quantities are defined in
Eq.~\eqref{eq:ed_rms_order_parameter}.  They provide a qualitative
finite-size comparison of the four ordering patterns; they are not
thermodynamic order parameters.  The four vertical lines are the gPEPS
phase boundaries inferred from Fig.~\ref{fig:order_params}.}
\label{fig:ed_op}
\end{figure}

The results are shown in Fig.~\ref{fig:ed_op}. The dominant order parameter changes across the phase diagram in the same
sequence as in the gPEPS calculation: AFM$_+$ near the
antiferromagnetic Heisenberg limit, AFM$_-$ between the AFM$_+$ region
and the Kitaev point at $\theta=\pi$, FM$_+$ near the ferromagnetic
Heisenberg limit, and FM$_-$ between the FM$_+$ region and the Kitaev
point at $\theta=0$.  In the Kitaev spin-liquid regimes, all four
order parameters are essentially suppressed compared to their values in the ordered regions.  This agreement supports the phase identification proposed in the main text.

\section{Majorana Hamiltonian for the Kitaev model and projective symmetry}
\label{app:free_majorana_hamiltonian}

We give the single-particle Majorana Bloch matrix in the ground-state flux sector of Sec.~\ref{sec:Kitaev_SHD}, using the gauge in Fig.~\ref{fig:kekule_coloring}.

Define the twelve-component spinor
\begin{equation}
\bm c_{\mathbf k}^{T}
=
\left(
c_{\mathbf k,A_0},\ldots,c_{\mathbf k,A_5},
c_{\mathbf k,B_0},\ldots,c_{\mathbf k,B_5}
\right),
\label{eq:appC_basis_order}
\end{equation}

where indices are shown in Fig.~\ref{fig:ed_36_cluster}(a). The momentum-space Hamiltonian and its  single-particle matrix are
\begin{equation}
H_{\rm K}^{\rm g.s.}
=
\frac14\sum_{\mathbf k\in{\rm BZ}}
\bm c_{-\mathbf k}^{T}\mathcal H(\mathbf k)\bm c_{\mathbf k}.
\label{eq:appC_H_iA}
\end{equation}
where, defining 
$z_i(\mathbf k)=\exp({\rm i}\mathbf k\cdot\mathbf n_i)$ for $i=1,2$, the 
Bloch matrix has the form 
\begin{widetext}
\begingroup
\setlength{\arraycolsep}{3pt}
\begin{equation}
\mathcal H(\mathbf k)
=
-\frac{{\rm i}}{2}
\left(
\begin{array}{cccccccccccc}
 0 &  K_x & 0 & 0 & 0 &  K_y & 0 & 0 & 0 & 0 & -K_z & 0 \\
 -K_x & 0 & -K_y & 0 & 0 & 0 & 0 & 0 & 0 & -K_z & 0 & 0 \\
 0 &  K_y & 0 &  K_x & 0 & 0 & K_z z_1 & 0 & 0 & 0 & 0 & 0 \\
 0 & 0 & -K_x & 0 & -K_y & 0 & 0 & 0 & 0 & 0 & 0 & K_z z_1 \\
 0 & 0 & 0 &  K_y & 0 &  K_x & 0 & 0 & K_z z_2 & 0 & 0 & 0 \\
 -K_y & 0 & 0 & 0 & -K_x & 0 & 0 & K_z z_2 & 0 & 0 & 0 & 0 \\
 0 & 0 & -K_z z_1^* & 0 & 0 & 0 & 0 & -K_y & 0 & 0 & 0 & -K_x \\
 0 & 0 & 0 & 0 & 0 & -K_z z_2^* & K_y & 0 & K_x & 0 & 0 & 0 \\
 0 & 0 & 0 & 0 & -K_z z_2^* & 0 & 0 & -K_x & 0 & -K_y & 0 & 0 \\
 0 & K_z & 0 & 0 & 0 & 0 & 0 & 0 & K_y & 0 & K_x & 0 \\
 K_z & 0 & 0 & 0 & 0 & 0 & 0 & 0 & 0 & -K_x & 0 & -K_y \\
 0 & 0 & 0 & -K_z z_1^* & 0 & 0 & K_x & 0 & 0 & 0 & K_y & 0
\end{array}
\right).
\label{eq:appC_bloch_matrix}
\end{equation}
\endgroup
\end{widetext}

The gauge-fixed matrix realizes the symmetry relations in
Eq.~\eqref{eq:projective_symmetry_relations}.  The matrices
$ U_{C_3}$, $ U_{C_2}$, and $ U_{M_x}$ below are
the explicit representatives of the corresponding sewing matrices in
that equation.  The matrix for the chiral  operator is
\begin{equation}
\Gamma = {\rm diag}(1,-1)\otimes{\rm diag}(1,-1,1,-1,1,-1),
\label{eq:appC_time_reversal}
\end{equation}

The threefold sewing matrix and its intra-hexagon permutation are
\begin{equation}
 U_{C_3}(\mathbf k)
=
\begin{pmatrix}
V & 0\\
0 & -{\rm e}^{{\rm i}\mathbf k\cdot\mathbf n_1}V
\end{pmatrix},
\qquad
V_{ij}
=
\delta_{i+2,j}.
\label{eq:appC_UC3}
\end{equation}
Here and below $i,j=0,\ldots,5$, with indices understood modulo six.

For the twofold rotation,
\begin{equation}
U_{C_2}=G_{C_2}\mathcal{P}_\sigma,\qquad G_{C_2}={\rm diag}(\mathds{1}_6,-\mathds{1}_6),
\label{eq:appC_UC2}
\end{equation}
Here, $\mathcal{P}_\sigma$  is the matrix associated with the index permutation $j\rightarrow \sigma(j)$, where \mbox{$\sigma=(1,10)(2,11)(3,12)(4,7)(5,8)(6,9)$}, with the identification
$j=1,\ldots,6,7,...,12\equiv A_0,\ldots,A_5,B_0,\ldots,B_5$.

The momentum-preserving antiunitary operator $\Theta$ acts as
$\Gamma U_{C_2}\mathcal K$ on the Bloch Hamiltonian $\mathcal{H}(\mathbf{k})$, and obeys
$\Theta^2=-\mathds{1}_{12}$, giving the doublets described in
Sec.~\ref{sec:projective_symmetry}.

For the mirror $M_x$ defined in Sec.~\ref{sec:projective_symmetry},
\begin{equation}
 U_{M_x}(\mathbf k)
=
\begin{pmatrix}
X_x & 0\\
0 & -{\rm e}^{{\rm i}\mathbf k\cdot\mathbf n_1}X_x
\end{pmatrix},
\qquad
(X_x)_{ij}
=
(-1)^i\,
\delta_{i+j,3}^{(6)}.
\label{eq:appC_UMx}
\end{equation}
Direct substitution verifies Eq.~\eqref{eq:M_projective} for arbitrary
$K_x,K_y,K_z$.

The matrices above realize the projective symmetries of
Eq.~\eqref{eq:projective_symmetry_relations} without enlarging the
twelve-site unit cell.

\section{Linear spin-wave theory}
\label{app:lswt}

This appendix gives the LSWT construction used in
Sec.~\ref{sec:phase_diagram}. 
We report LSWT energies only where the associated quadratic bosonic  Hamiltonian is stable.  LSWT is not controlled in the Kitaev spin-liquid regions because there is no ordered
reference state.

For any of the collinear orders \eqref{eq:classical_collinear_ansatz}, introduce Holstein--Primakoff bosons
$a_{\mathbf R,a}$ for the spin operators
\begin{subequations}
\label{eq:lswt_HP}
\begin{align}
\mathbf S_{\mathbf R,a}\cdot \mathbf a_a
&=
\sqrt{\frac{S}{2}}
\left(
a_{\mathbf R,a}
+
a_{\mathbf R,a}^{\dagger}
\right)
+O(S^{-1/2}),
\label{eq:lswt_HP_x}
\\
\mathbf S_{\mathbf R,a}\cdot \mathbf b_a
&=
\sqrt{\frac{S}{2}}
\frac{
a_{\mathbf R,a}
-
a_{\mathbf R,a}^{\dagger}
}{i}
+O(S^{-1/2}),
\label{eq:lswt_HP_y}
\\
\mathbf S_{\mathbf R,a}\cdot \mathbf c_a
&=
S
-
a_{\mathbf R,a}^{\dagger}a_{\mathbf R,a}.
\label{eq:lswt_HP_z}
\end{align}
\end{subequations}
where $\mathbf{R}$ labels the unit cell and $a$ labels sites in the unit cell. For all four LSWT calculations reported here, we choose
$\hat{\mathbf n}=\hat{\mathbf z}$ and set
$\mathbf c_a=f_a^\mu\hat{\mathbf z}$; $\mathbf a_a$ and
$\mathbf b_a$ complete a right-handed orthonormal triad.
For AFM$_+$ and FM$_+$, the classical energy is accidentally
independent of $\hat{\mathbf n}$ even though the Kitaev interaction
breaks $SU(2)$ symmetry. The spin-wave spectra reported below
correspond specifically to the choice $\hat{\mathbf n}=\hat{\mathbf z}$.

The bosonic operators \eqref{eq:lswt_HP} convert the isotropic Kitaev-Heisenberg Hamiltonian into the quadratic LSWT Hamiltonian \begin{equation}
H_{\rm LSWT}
=
E_{\rm cl}
+
\frac{1}{2}
\sum_{\mathbf q\in{\rm BZ}}
\left[
\Phi_{\mathbf q}^{\dagger}
{\cal H}_{\rm BdG}(\mathbf q)
\Phi_{\mathbf q}
-
{\rm Tr}\,A(\mathbf q)
\right].
\label{eq:lswt_bdg_hamiltonian}
\end{equation}
where we defined the Nambu spinor
\begin{equation}
\Phi^\dag_{\mathbf q}
=
\left(
a^\dag_{\mathbf q,A_0},...,
a^\dag_{\mathbf q,B_5},
a_{-\mathbf q,A_0},...,
a_{-\mathbf q,B_5}\right),
\label{eq:lswt_nambu_spinor}
\end{equation}
here $a_{\mathbf{q},a}$ is the Fourier transform of $a_{\mathbf{R},a}$. The bosonic BdG matrix is
\begin{equation}
{\cal H}_{\rm BdG}(\mathbf q) =
\begin{pmatrix}
{\mathcal A}(\mathbf q) & {\mathcal B}(\mathbf q)\\
{\mathcal B}^\dagger(\mathbf q) & {\mathcal A}^T(-\mathbf q)
\end{pmatrix},
\label{eq:lswt_bdg_matrix}
\end{equation}
where ${\mathcal A}(\mathbf q)$ is the normal hopping block and ${\mathcal B}(\mathbf q)$ is the anomalous pairing block.  Both are $12\times12$ matrices. The blocks $A$ and $B$ are assembled bond by bond.

Next, we diagonalize the matrix $\Sigma_z{\cal H}_{\rm BdG}(\mathbf q)$,   where \mbox{$\Sigma_z
=\mathrm{diag}(\mathbbm{1}_{12}, -\mathbbm{1}_{12})$}. The eigenvalues of $\Sigma_z{\cal H}_{\rm BdG}(\mathbf q)$ occur in positive and negative particle-hole pairs $\pm \omega_\nu$. Energetic stability is tested directly from the eigenvalues of ${\cal H}_{\rm BdG}$, as we detail below.

The zero-point corrected LSWT
ground-state energy is
\begin{equation}
E_{\rm LSWT}
=
E_{\rm cl}
+
\frac{1}{2}
\sum_{\mathbf q\in{\rm BZ}}
\left[
\sum_{\nu=1}^{12}
\omega_\nu(\mathbf q)
-
{\rm Tr}\,{\mathcal A}(\mathbf q)
\right],
\label{eq:lswt_energy}
\end{equation}
where the sum is performed over the first Brillouin-zone. Here, we discretized this zone using a  $18\times18$ mesh.  We checked that increasing the mesh to $24\times24$ and \mbox{$30\times30$} changes $E_{\rm LSWT}/N$ by less than $10^{-6}$. For instance, at the antiferromagnetic Heisenberg point $(K,J)=(0,1)$, previous studies~\cite{Richter04} gives
$E_{\rm LSWT}/N=-0.538908$, for a $18\times18$ mesh, whereas we obtain $E_{\rm LSWT}/N=-0.538909$, for the $30\times30$ mesh.

\begin{figure}[t]
\centering
\includegraphics[width=\columnwidth]{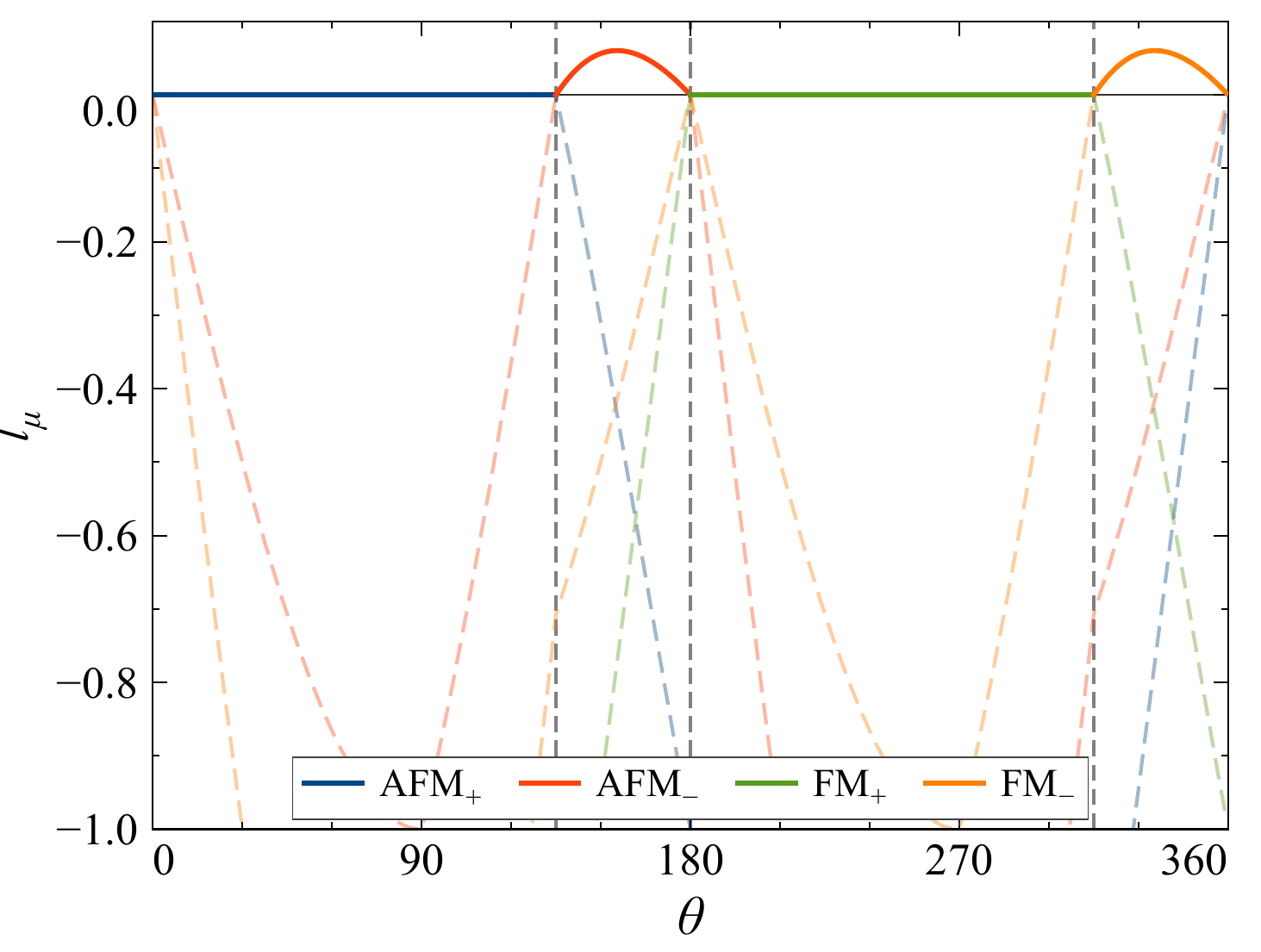}
\caption{Smallest eigenvalue $\l_\mu(\theta)$ of $\mathcal{H}_{\rm BdG}(\mathbf q)$ for the four collinear states.  Solid portions denote the
classical intervals in Table~\ref{tab:classical_gs}; dashed extensions
locate the loss of stability. Vertical lines mark the classical boundaries.}
\label{fig:lswt_stability}
\end{figure}

For a given magnetic order $\mu$ [see Eqs.~\eqref{eq:f_AFMp}-\eqref{eq:f_FMm}], energetic stability requires the Hermitian quadratic form to be positive semidefinite. We monitor $l_\mu(\theta)$, defined as the smallest eigenvalue of $\mathcal{H}_{\rm BdG}(\mathbf q)$ in the first Brillouin zone for a fixed magnetic order $\mu$ and a fixed $\theta$. We retain a branch only when $\l_\mu(\theta)\geqslant0$.  As shown in Fig.~\ref{fig:lswt_stability}, the stable intervals agree, within this
resolution, with Table~\ref{tab:classical_gs}.  
The near-zero values for AFM$_+$ and FM$_+$ reflect the accidental continuous degeneracy of the corresponding classical spin orientations.

\begin{figure}[t]
\centering
\includegraphics[width=\columnwidth]{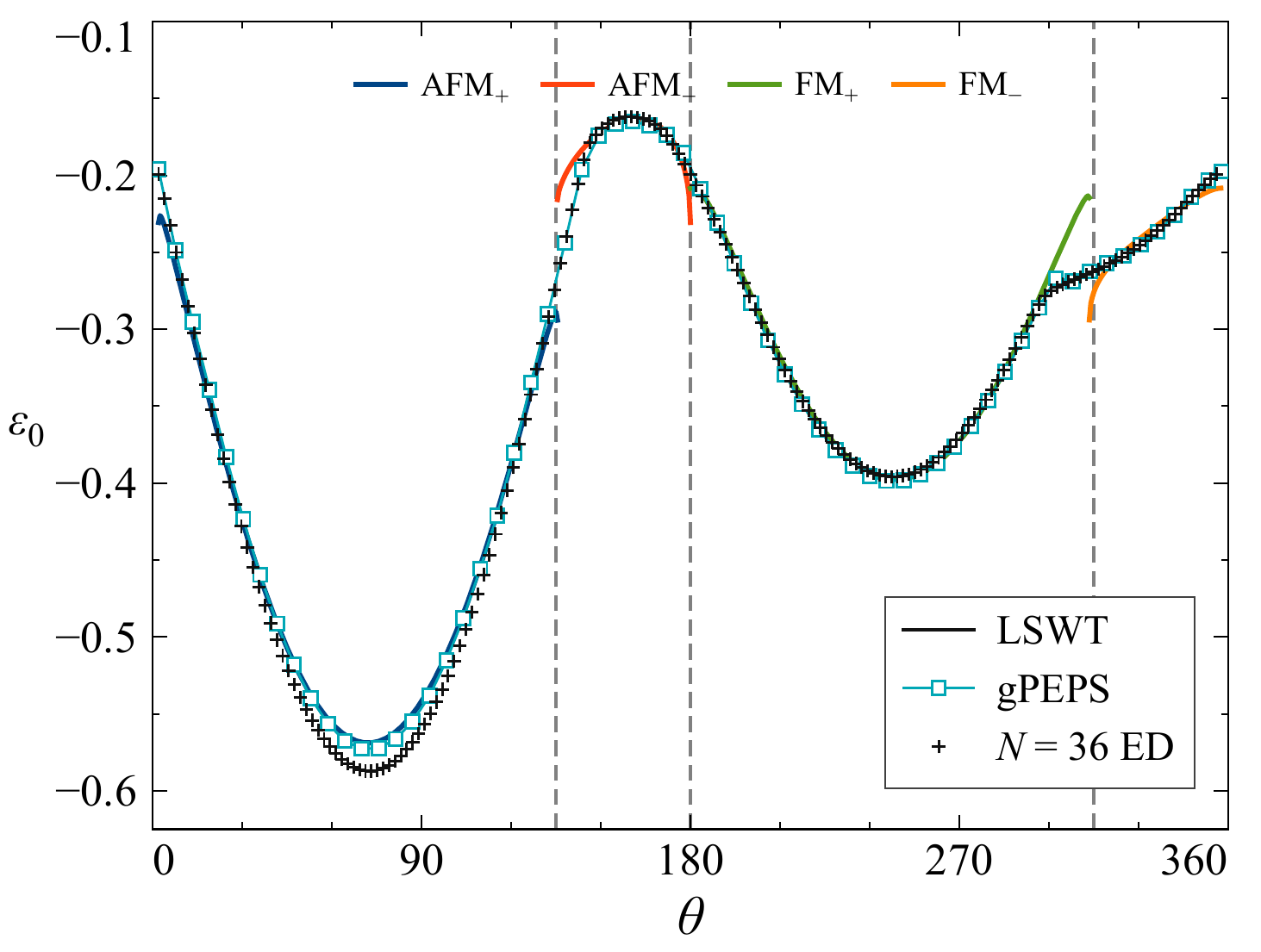}
\caption{Stable LSWT energies compared with gPEPS and the lowest
$N=36$ ground state energy per site $\varepsilon_0$ in the fully symmetric irrep.  Each LSWT branch is shown
only on its classical stability interval; vertical lines mark the classical boundaries. The black star marks the exact product-state benchmark at $\theta_\star=\pi-\tan^{-1}(1/2)$, where $\varepsilon_0=-3/(8\sqrt5)$.}
\label{fig:lswt_stable_energy}
\end{figure}

The energy per site, $E_{\rm LSWT}/N$, is shown in
Fig.~\ref{fig:lswt_stable_energy} along with the ground-state energy per site computed from ED and gPEPS, replotted from Fig.~\ref{fig:energy_kh}. The three curves show a good agreement inside the ordered phases and large corrections near the Kitaev limits.  The offsets between one-sided branches at a classical
boundary are not physical discontinuities: the competing states have equal classical energies there but generally different harmonic zero-point corrections.  The LSWT is therefore used only as a phase-interior benchmark; the phase boundaries being inferred from gPEPS and supported by ED. Note that at the exact solvable point $(K,J) = (-2,1)$, the AFM$_-$ reference state is the exact product ground state. Consequently, the LSWT zero-point correction vanishes and
$E_{\rm LSWT}/N=-3/8$, which is the classical ground state energy at this parameter point.

\section{Ground state and hidden $U(1)$ symmetry at $(K,J) = (-2,1)$}\label{app:exact_solvable_point}

In this appendix we provide an alternative derivation for the ground state of the Kitaev--Heisenberg model at the point $(K,J) = (-2,1)$. We perform a unitary transformation
\begin{equation}
U H U^\dag = -\sum_{\langle ij\rangle}
\left(S^z_iS^z_j + \epsilon_{ij}(S^x_iS^x_j+S^y_iS^y_j)\right) \label{xxze}
\end{equation}
where  $\epsilon_{ij} := \left\{\begin{array}{ll} -1& x\text{ bond},\\
+1 & y \text{ or }z\text{ bond},\end{array}\right.$
and the unitary transformation $U$ is defined via
\begin{equation}
U = \prod_{\mathbf{R}}
{\rm e}^{{\rm i} \pi \left(\sum_{ i=0,2,4} \hat{S}^z_{\mathbf{R}, A_i} + 
\sum_{ i=1,3,5} \hat{S}^x_{\mathbf{R}, A_i} 
+
\sum_{ i=1,3,5} \hat{S}^y_{\mathbf{R}, B_i}\right)}
\end{equation}
where $\mathbf{R}$ labels the unit cell, and $A_i$, $B_i$ labels sites inside the unit cell as shown in Fig.~\ref{fig:ed_36_cluster}(a).


The transformed Hamiltonian $\widetilde{H} := U H U^\dag$, defined in Eq.~\eqref{xxze}
has a simple structure: it consists of ferromagnetic Heisenberg exchanges on the $z$ bonds, and alternating mixed-sign exchanges on the ($x$ and $y$) bonds of the hexagons $\hexagon$. 

Define $H_{\hexagon}$ as the ``Hamiltonian'' associated with a single hexagon $\hexagon$. One can easily show that its ground states are the maximally polarized states 
\begin{equation}
|\!\Uparrow\rangle :=|\!\uparrow\uparrow\uparrow\uparrow\uparrow\uparrow\rangle \text{ and }
	|\!\Downarrow\rangle := |\!\downarrow\downarrow\downarrow\downarrow\downarrow\downarrow\rangle,
\end{equation}
then the global ground states of $\widetilde{H}$ can be assembled from maximally aligning the states $|\!\Uparrow\rangle$ and $|\!\Downarrow\rangle$ on each hexagon, which gives $U|\Psi_{\text{AFM}_-}\rangle$ in Eq.~\eqref{eq:exact_AFMm_state} and its time-reversal partner $\mathcal{T}U|\Psi_{\text{AFM}_-}\rangle$. This construction makes it clear that these are the only ground states, as no other states can simultaneously minimize $H_{\hexagon}$ and the $z$ bonds.




This construction also identifies a $U(1)$ symmetry of the Hamiltonian:
\begin{equation}
[\widetilde{H}, \widetilde{\mathcal{S}}^z]=0,~~~
\widetilde{\mathcal{S}}^z:=
\sum_i S^z_i.
\end{equation}
In the original basis of $H$, the conserved quantity writes as \begin{equation}
\mathcal{S}^z:= U^\dag\widetilde{\mathcal{S}}^z U =
\sum_{i} f_i^{{\rm AFM}_-} S^z_i,
\end{equation}
which defines a staggered magnetization. This conserved staggered magnetization generates the hidden $U(1)$ symmetry discussed in Sec.~\ref{sec:exact_product_state}. This is in contrast to the Kitaev model on the \emph{honeycomb lattice} with the usual Kitaev coloring, where we have a hidden SU(2) symmetry \cite{Chaloupka15}.

%


\end{document}